\documentclass[a4paper,fleqn,usenatbib]{mnras}

\newcommand{\atl}{ATLAS$^{\rm 3D}$}

\newcommand{\re}{\hbox{$R_{\rm e}$}}
\newcommand{\remaj}{\hbox{$R_{\rm e}^{\rm Maj}$}}
\newcommand{\mlam}{(M$_*$/L)$_{dyn}^{AM}$~}
\newcommand{\mljam}{(M/L)$_{JAM}$}
\newcommand{\mlpl}{(M$_*$/L)$_{dyn}^{PL}$~}
\newcommand{\mlsalp}{(M$_*$/L)$_{pop}^{Salp}$~}

\usepackage{longtable}
\usepackage{setspace}
\usepackage{amsmath}
\usepackage{amssymb}
\usepackage{amsmath}
\usepackage[flushleft]{threeparttable}
\usepackage{booktabs,caption}
\usepackage{color}
\usepackage{textcomp}
\usepackage{lscape}
\usepackage{geometry}
\usepackage{floatpag}
\usepackage{graphicx}
\usepackage{natbib}
\usepackage{color}
\usepackage{gensymb}
\usepackage{afterpage}
\usepackage[11pt]{moresize}
\usepackage{afterpage,lscape}
\usepackage{float}

\usepackage{hyperref}
\hypersetup{
	colorlinks=true,
	linkcolor=blue,
	filecolor=magenta,      
	urlcolor=cyan,
}

\title[A Precise Benchmark for Cluster Scaling Relations]{A Precise Benchmark for Cluster Scaling Relations: Fundamental Plane, Mass Plane and IMF in the Coma Cluster from Dynamical Models}

\author[S. Shetty et al.]{
Shravan Shetty,$^{1}$\thanks{E-mail: shravan.shetty@pku.edu.cn}
Michele Cappellari,$^{2}$ Richard M. McDermid,$^{3}$ Davor Krajnovi\'{c},$^{4}$ 
\newauthor P. T. de Zeeuw$^{5,6}$, Roger L. Davies$^{2}$ and Chiaki Kobayashi$^7$
\\
\\
$^{1}$Kavli Institute for Astronomy and Astrophysics, Peking University, Beijing 100871, China\\
$^{2}$Sub-Department of Astrophysics, Department of Physics, University of Oxford, Denys Wilkinson Building, \\ Keble Road, Oxford OX1 3RH, UK\\
$^{3}$Department of Physics and Astronomy, Macquarie University, Sydney, NSW 2109, Australia \\
$^{4}$Leibniz-Institut f\"{u}r Astrophysik Potsdam (AIP), An der Sternwarte 16, D-14482 Potsdam, Germany \\
$^{5}$Sterrewacht Leiden, Leiden University, Postbus 9513, 2300, RA Leiden, The Netherlands \\ 
$^{6}$Max-Planck-Institut f\"{u}r Extraterrestrische Physik, Giessenbachstra$\beta$e, 85741, Germany \\
$^{7}$Centre for Astrophysics Research, University of Hertfordshire, College Lane, Hatfield AL10 9AB, UK \\
}

\date{Accepted XXX. Received YYY; in original form ZZZ}

\pubyear{2019}

\begin{document}
\label{firstpage}
\pagerange{\pageref{firstpage}--\pageref{lastpage}}
\maketitle

\defcitealias{atlas3d15}{A$^{\rm 3D}$15}
\defcitealias{atlas3d20}{A$^{\rm 3D}$20}
\defcitealias{Cappellari2016_review}{C16}

\begin{abstract}

We study a sample of 148 early-type galaxies in the Coma cluster using SDSS photometry and spectra, and calibrate our results using detailed dynamical models for a subset of these galaxies, to create a precise benchmark for dynamical scaling relations in high-density environments. For these galaxies, we successfully measured global galaxy properties, modeled stellar populations, and created dynamical models, and support the results using detailed dynamical models of 16 galaxies, including the two most massive cluster galaxies, using data taken with the SAURON IFU. By design, the study provides minimal scatter in derived scaling relations due to the small uncertainty in the relative distances of galaxies compared to the cluster distance. Our results demonstrate low ($\leq$55\% for 90$^{th}$ percentile) dark matter fractions in the inner 1\re ~of galaxies. Owing to the study design, we produce the tightest, to our knowledge, IMF-$\sigma_e$ relation of galaxies, with a slope consistent with that seen in local galaxies. Leveraging our dynamical models, we transform the classical Fundamental Plane of the galaxies to the Mass Plane. We find that the coefficients of the mass plane are close to predictions from the virial theorem, and have significantly lower scatter compared to the Fundamental plane. We show that Coma galaxies occupy similar locations in the (M$_*$ - \re) and (M$_*$ - $\sigma_e$) relations as local field galaxies but are older. This, and the fact we find only three slow rotators in the cluster, is consistent with the scenario of hierarchical galaxy formation and expectations of the kinematic morphology-density relation.

\end{abstract}

\begin{keywords}
galaxies:clusters:general -- galaxies:evolution -- galaxies:kinematics and dynamics -- galaxies:fundamental parameters.
\end{keywords}

\section{Introduction}

With the onset of Integral-Field Spectrographic surveys of galaxies, significant advances have been made in our understanding of the properties of galaxies as a class (see \citealt{Cappellari2016_review} [hereafter \citetalias{Cappellari2016_review}] for a review). Initial surveys such as the SAURON survey \citep{SAURONsurvey} and the \atl ~~survey \citep{atlas3d1} studied the properties of hundreds of early-type galaxies (ETGs), while the DISKMASS survey \citep{DISKMASS} investigated spiral galaxies. These surveys have since then been followed up by other surveys such as CALIFA \citep{CALIFA} that observed both spiral and early-type galaxies using a sample size of \textasciitilde{600}. Today ever-larger surveys such as the SAMI \citep{Bryantetal2015} and MaNGA \citep{MANGA} are currently underway with the aim of extending the sample of galaxies observed using Integral-Field Units (IFU) to thousands of galaxies in the local universe. Simultaneously, other IFU surveys are ongoing to further develop our understanding of individual galaxy classes. Such examples include the MASSIVE survey \citep{MASSIVE} which aims to study massive ETGs out to twice their effective radii, M3G survey \citet{Krajnovicetal2018} that has observed massive galaxies at the densest environments, and the recent Fornax3D project that aims to trace the evolution of ETGs, and the potential transformation of these galaxies into late-type galaxies, in dense environments \citep{Fornax_MUSE}. These surveys allow astronomers to study the resolved properties of the galaxies in the local universe within a statistical framework and study their variations with global galaxy properties such as morphology, assembly history, mass, etc.

For several years a great effort has been devoted to study the scaling relations of galaxies, such as the Faber-Jackson relationship \citep{faberjackson1976}, Tully-Fisher relationship \citep{Tully-Fisher}, and Fundamental Plane \citep{djorgovski1987, dressler1987}. This has been particularly true of the Fundamental Plane, which relates the luminosity of the galaxy to its size (\re) and kinematics ($\sigma$). The Fundamental Plane can be interpreted as being an outcome of the galaxies being in virial equilibrium, however observations of the Fundamental Plane demonstrate a clear deviation of the relation from that expected if light follows mass, $L \propto \sigma^{2} R_{\rm e}$ \citep{Jorgensenetal1996,Hudsonetal1997,Scodeggioetal1998,Pahreetal1998, Collessetal2001, Gibbonsetal2001, Bernardietal2003, HydeBernardi2009, LaBarberaetal2010, Magoulasetal2012}. This deviation of the coefficients has been termed the ``tilt'' of the Fundamental Plane. It has been suspected that the observed tilt is due to the approximations required to translate the virial equation to the observable quantities; particularly assumptions of a constant mass-to-light ratio (M/L) of the stellar population of the galaxies, and that of homology in the structure of the galaxies \citep{RenziniCiotti1993, GrahamColless1997, PrugnielSimien1996, Forbesetal1998, Bertinetal2002,Tortoraetal2012}. Studies by \citet{cappellari2006}, \citet{atlas3d15} (hereafter \citetalias{atlas3d15}), ~\citet{Boltonetal2007}, and \citet{auger2010b} have used dynamical or strong lensing techniques to demonstrate that the tilt in the Fundamental plane can be explained by a systematically varying IMF and/or the Dark Matter fraction within galaxies. 

Recent results such as those by \citet{vandokkum2010}, \citet{vanDokkum2011} etc have hinted at the non-universality of the IMF. Studies based on the analysis of stellar dynamics \citep{cappellari2012nature,atlas3d20,Posackietal2015_SALCS_IMF,Lietal2017}, and those based on an analysis of stellar populations \citep{Conroy2012,Spiniello2012,Smith2012,Ferreras2013,LaBarbera2013} demonstrate that the IMF normalization of the galaxies varies systematically with the velocity dispersion of the galaxies. Building upon these results, \citet{atlas3d20} (hereafter \citetalias{atlas3d20}) demonstrate that accounting for this non-universality transforms the observed Fundamental Plane, with the ``tilt'', into the virial condition and hence non-edge-on projection of the mass plane contains the information on the evolution of the galaxies. However despite the qualitative consistency between the independent analyses there is a quantitative inconsistency between the results \citep{Smith2014,ATLAS3D30}. Furthermore, several publications have presented evidence suggesting that the variation in the IMF of the galaxies may be due to systematic changes in the metallicities of galaxies \citep{MartinNavarroetal2015_CALIFA,Zhouetal2019}, and the observed dependence with velocity dispersion is due to the correlation between the velocity dispersion and metallicity in galaxies.

In this study, we explore the galaxies of the Coma cluster using the state-of-the-art dynamical modelling techniques to study the IMF--$\sigma$ relation of the galaxies and their scaling relations. We use models to study the IMF-$\sigma$ relation of galaxies and their scaling relations. By studying the Coma cluster, we aim to address the most significant source of scatter in these relations, the uncertainty in the relative distance to the galaxies, and hence minimize the scatter in the relations. While the Coma cluster is close enough to be feasibly observed using an Integral-Field Spectrograph (IFS), the cluster is far and dense enough to reduce any uncertainty of their relative distance to less than 2\%. Using a sample of 53 galaxies observed using the SAURON IFS \citep{SAURON}, including the two Brightest Cluster Galaxies (BCGs), and a larger sample of 148 galaxies with SDSS photometric and spectral data \citep{SDSS}, we created robust dynamical and stellar population models for the galaxies to study their scaling relations and compare these to those observed in the local low-density environments. 

In the following Section~2, we present a brief description of the selection criteria for the studied samples. In Section~3 and 4, we describe the data used and the methodology of our analysis. Section~5 discusses the results of this study, and in Section~6 we present a summary of our results. For this study we assume that the Coma cluster is at a distance of 100Mpc and the universe is flat with $\Omega_{m} = 0.3$, $\Omega_{\Lambda} = 0.7$ and $H_{0}=70$kms$^{-1}$Mpc$^{-1}$.

\section{Galaxy Samples}

The galaxy sample for this study is selected from the sample of 161 galaxies used in \citet{Cappellari2013ApJL}, where the author selected galaxies within one square degree of the Coma cluster center with total K-band absolute magnitude limit of M$_K < -21.5$, which translates to $M_* \gtrsim 6 ~\times 10^9 ~M_{\odot}$, at a distance of 100Mpc. These criteria were designed to be identical to the \atl~ survey {\em parent} sample, from which the ETGs of the \atl ~~survey were extracted. Here we use this catalog as our parent sample so that the results can be directly compared to that of \atl. 

Of this sample of 161 galaxies, we selected 148 galaxies based on the availability of spectral data in SDSS 12$^{th}$ data release catalog \citep{SDSS_DR12}. For a subset of 53 galaxies, we have data taken using the SAURON IFS mounted on the William Herschel Telescope. These 53 galaxies inhabit the central region of the cluster and include the two BCGs of the Coma cluster. Of these 53 galaxies, due to issues of Signal-to-Noise (S/N), unexpected artifacts in the data, etc., dynamical models for only 42 galaxies could be made. Of these, two galaxies did not have SDSS spectral observations. For simplicity of presentation, we ignore this fact in what follows we still refer to the galaxies with SAURON data as a subset of the main sample.

\begin{figure}
	\centering
	\includegraphics[width=\columnwidth]{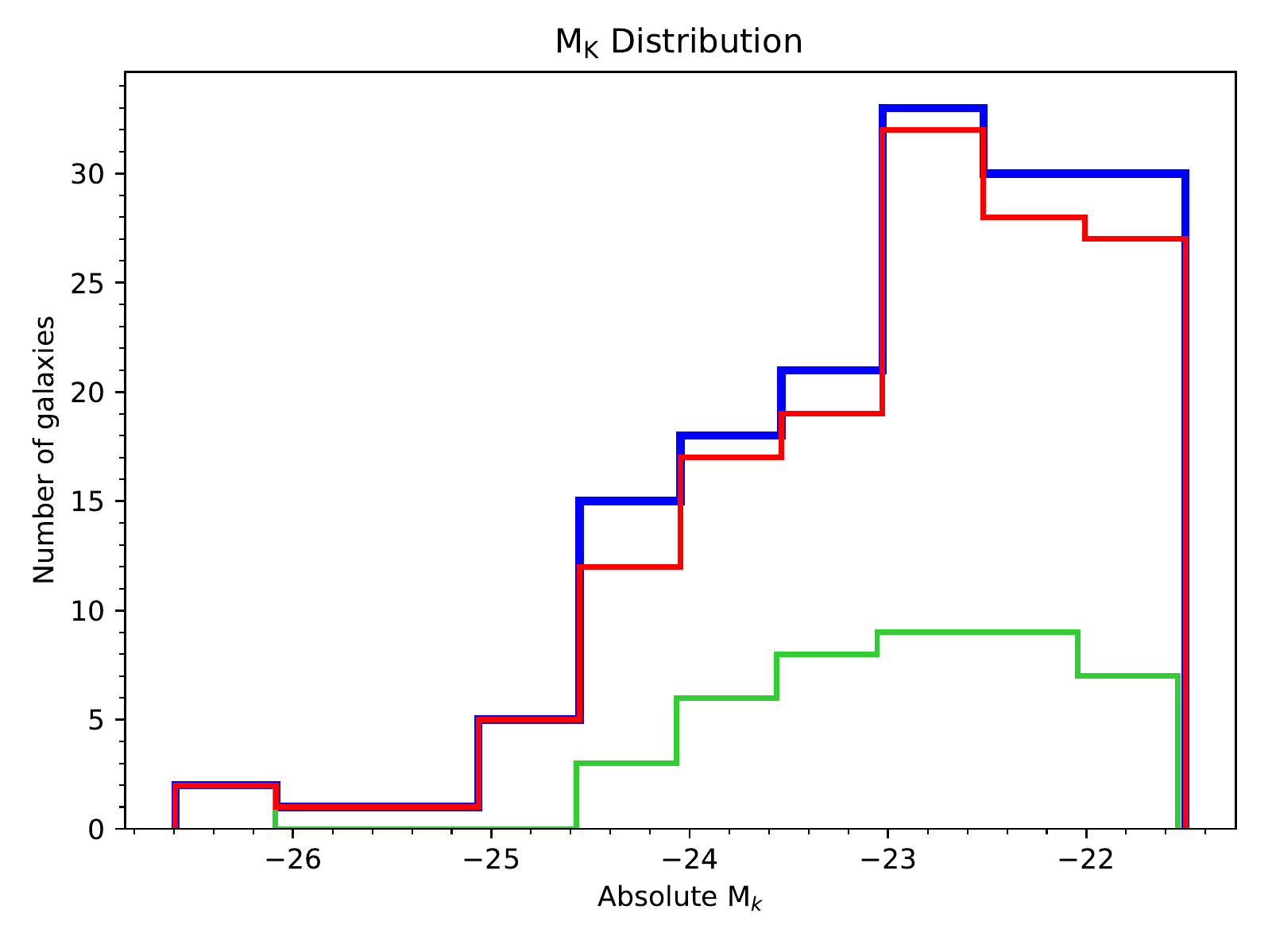}
	\caption[Sample Comparison]{Here we present the histogram distribution of the original sample of 161 galaxies in blue, and compare it to the two sample sets used in this study; the full sample represented in red and the SAURON subset represented by green. The histogram demonstrates that both samples represent the original in an unbiased manner, allowing the results of this study to be applicable to the cluster in general.}
	\label{sample_KS_test}
\end{figure}

In Fig.~\ref{sample_KS_test}, we present the M$_{\rm K}$ distribution of the full sample and the SAURON subset and compare them to the original sample of 161 galaxies. The histogram demonstrates that the full sample, and the SAURON subset, are representative of the observed distribution of the M$_{\rm K}$ of the original sample. To quantify this, we conducted the two-sample Kolmogorov-Smironov test \citep[Chapter.~7 of][]{pratt1981kolmogorov}, finding that the null hypothesis, i.e., two samples are derived from the same parent sample, is accepted with p-values of 1 and 0.99 for the full sample and SAURON subset respectively. The unusually high p-values for these tests are likely because p-values are random variables that are biased to lower values when the null hypothesis is false. Hence a high value of p-value is only indicative of the acceptance of the hypothesis but not a quantification of it \citep{bland2013baseline}.

\section{Data}

\subsection{SDSS Photometric and Spectroscopic Data}

The generation of dynamical models for observed stellar kinematics requires one to specify (i) the distribution of the total mass, from which the gravitational potential is generated, and (ii) the distribution of the ``tracer'' population, for which one wants to predict the kinematics with the models. The latter is typically obtained by deprojecting the surface brightness of the galaxies. In this study, we use the $r$-band mosaic images provided by SDSS \citep{SDSS} in their 12$^{th}$ data release \citep{SDSS_DR12} to do this, due to the high S/N of the filter. Additionally, this band is close in wavelength coverage to the wavelength range used to derive the stellar kinematics of the full sample (as will be described in Section.~\ref{full_sample_kinematics}) and hence closely describes the ``tracer'' population of the stellar kinematics. The median $r$-band point spread function (PSF) width of the survey is 1.43$''$, and the images used in this study have a spatial scale of 0.396$''$/pixel.

In addition to the photometric data, we also used spectral data from SDSS for the 148 galaxies in the full sample. While most galaxies were observed as part of the SDSS-I/II survey \citep{SDSS}, two galaxies were observed in the BOSS survey in SDSS-III \citep{Dawsonetal2013_SDSS_BOSS}. Consequently, the data were observed either using the original SDSS spectrographs, or the BOSS spectrograph. These instruments observe using different aperture diameters, i.e., 3$''$ and 2$''$ respectively, which is vital to note when modelling the dynamics of the galaxies. Both these instruments observe in similar wavelength ranges, around 3,800~--~9,200\AA, and at similar spectral resolution (\textasciitilde{69}km/s) though the BOSS instrument does observe out to slightly higher and lower wavelengths. For a detailed comparison of the two instruments, we refer the interested reader to \citet{Ahnetal2012_SDSS_BOSS}.

\subsection{SAURON IFS data}

In addition to the SDSS data, we have obtained IFS data for 53 galaxies using the SAURON instrument mounted on the William Herschel Telescope in the La Palma Observatory. The SAURON instrument was a lenslet-based IFS, consisting of \textasciitilde{1,400} spaxels, capable of observing in the wavelength range 4,810~--~5,350\AA~ \citep{SAURON}. We observed our galaxies in the low-resolution mode of SAURON, which observes $41'' \times 33''$ of the sky at a spatial resolution of $0.94''$/spaxel. The spectral resolution of the instrument is 3.9\AA ~FWHM, which corresponds to an instrumental velocity dispersion of 98 km s$^{-1}$. The galaxies were observed in March and May 2012 and April 2013 with a 1-hour exposure on each galaxy split between two dithered 30-minute exposures.

The data were reduced using the \textsc{XSAURON} software \citep{Baconetal2001_SAURON1, Emsellemetal2004_SAURON3}, which implements bias subtraction, flat-fielding, wavelength calibration, and sky-subtraction on the observed data to produce the final ``data cubes'', where each $(x,y)$ coordinate in the file has a corresponding spectrum. 

\subsection{High-Resolution Photometry}

In order to create accurate dynamical models for of the two BCGs, and hence measure their Supermassive Black Holes (SMBH), we require high-resolution photometry to generate high-resolution mass maps. For this we utilize \textit{Hubble Space Telescope} photometry observed as part of the \textit{HST} program \textit{GO}--11711 by J. Blakeslee. These images were taken with the \textit{Advanced Camera for Surveys Wide Field Channel (ACS/WFC)} using the filter F814W. The observations were dithered to improve bad pixel rejection and analyzed using the standard pipeline of the Space Telescope Science Institute's Mikulski Archive for Space Telescopes (MAST). The final image has a resolution of 0.05''/pixel and a PSF of 0.1-0.14'' FWHM, as modeled by the \textsc{TinyTim} software package \citep{TinyTim}.

\section{Analysis}

\subsection{Stellar Kinematics}

To model the observed spectra of galaxies, we used the penalized pixel fitting (\textsc{pPXF}) method and code by \citet{ppxf} and updated by \citet{ppxf_2}\footnote{\label{noteMCsoftware} http://purl.org/cappellari/software}. This code fits the observed galaxy spectra with a set of template spectra in pixel space using a Gaussian parametrization for the line-of-sight velocity distribution (LOSVD) of the galaxy \citep{Gerhard1993,vanderMarelFranx1993}. In addition to this, \textsc{pPXF} also permits the use of multiplicative and additive polynomials. These polynomials can account for some variation between the continuum shape, which can arise due to the galaxy spectrum and template spectrum being observed by different spectrographs, and template mismatch.

We use the Indo-US Library of Coud\'{e} Feed Stellar Spectra Library \citep{indo-uslibrary} as our template set. This library consists of 1,273 stellar spectra covering the wavelength range of 3,460--9,464\AA~ at a resolution of \textasciitilde{1.35}$\AA$~FWHM \citep{Beifiori2011}. This template set was selected as it has the most extensive coverage of stellar parameters at a resolution higher than that of the SDSS spectra within the wavelength range used to derive the stellar kinematics (See Section.~4.1.1). Also, we assume the LOSVD of our galaxies to have a gaussian shape, i.e., we fit for only the stellar velocities and dispersions of the LOSVD and set the higher-moments of kinematics to zero.

\subsubsection{Kinematics for the full Sample}
\label{full_sample_kinematics}

The velocity and the velocity dispersion of each galaxy was derived using the SDSS spectra of the galaxies. These spectra cover a large range in wavelength, however, in this study, we compare and calibrate the analysis of the full sample with that of the SAURON subset and hence to ensure comparable analysis we use only the wavelength range of the SAURON IFU, i.e., 4,810~--~5,350\AA, to derive the kinematics of the full sample. 

\subsubsection{Kinematics for the SAURON Subset}
\label{sauron_kin}

To fit for the individual spectra of the SAURON datacubes, we use the same technique as described above. However, due to the light profile of galaxies, the S/N in each spaxel of an IFS data cube spans a large range of values. Hence, when working with IFS data it is common practice to spatially bin the results over the plane of the IFS so as to recover information within spaxels with low S/N. For the SAURON data, we bin the low S/N spaxels using the \textsc{voronoi\_2d\_binning} code\textsuperscript{\ref{noteMCsoftware}} \citep{CappellariCopin2003_Voronoi_binning} which bins the spaxels according to Voronoi tessellations \citep{Okaeetal_tesellations}. The code does so while not strictly forcing the solution to the desired S/N as this may result in the loss of spatial information. 

In order to bin spaxels to a desired S/N, we first need to derive reliable estimates for the S/N of the spaxels. The data reduction software, \textsc{XSAURON}, does include within the data cube the formal error for each spectral pixel. Thus a formal estimate of the S/N of the spaxels is trivial. However, to test the reliability of the formal estimate, we compared it with S/N derived using the residuals from the full spectrum fitting of the spectra. This test demonstrated a linear relationship between the two quantities however the formal estimate did over-state the S/N compared to the latter. Hence before binning, the S/N of the spaxels was corrected for this over-estimation.

Based on these calibrated noise estimates, we clip spaxels with S/N less than 1, as these spaxels are noise-dominated, and then use \textsc{voronoi\_2d\_binning} to bin the spaxels to a target S/N of 25. At this value, the kinematic maps generated are consistent with those generated at higher values, while retaining the highest spatial information of the data. 

Template mismatch can have severe consequences on the derived kinematics of galaxies. Hence, it is crucial to take steps to minimize the effect that this may have when deriving the kinematics for bins with low S/N. To make our results robust to this effect when fitting low S/N spaxels, for each galaxy we derive an optimal template by fitting the high-quality spectrum derived by co-adding the spaxels within the effective radii of the galaxy. This optimal template is then used to fit for the kinematics of all the bins within the data cube. In these fits, we use additive Legendre polynomials of degree 4 to account for any variations in the stellar populations across the field of view of the IFU. To test the efficacy of this approach, we compared the measured kinematics with that measured by fitting each bin with the stellar template set and find no significant systematics between the two quantities.

\subsection{Parametrization of Surface Brightness}
\label{surf_brightness_parametrization}

As mentioned previously, the dynamical modelling of galaxies requires a parametrization of the galaxy surface brightness. We did this using the Multi-Gaussian Expansion (MGE) parametrization \citep{emsellemetal1994,cappellari2002} as implemented by the \textsc{mge\_fit\_sectors} code\textsuperscript{\ref{noteMCsoftware}}. This technique can fit for multiple components of a galaxy while allowing for an analytical accounting for the point spread function of the instrument and deprojection of the observed light when the inclination is known. 

For the galaxies in the Coma cluster, we modeled the surface brightness of the galaxies using their \textit{r}-band mosaic images obtained from SDSS DR12 \citep{SDSS_DR12}.  The images were downloaded as thumbnails of size $75''\times75''$ for most galaxies, while for smaller galaxies these thumbnails were reduced to $40''\times40''$.  Visual observations of these thumbnails verified the complete coverage of the galaxy light with additional sky coverage to estimate the sky and noise levels. 

When fitting for the surface brightness of galaxies, it is critical to ensure that the pixels used for the procedure contain information of the intrinsic light profile of the galaxies. During the parametrization process, we realized that light halos, caused by the internal reflection of the galaxy light within the telescope's optical components \citep{Slater_harding_Miho_2009_Internal_Reflections}, had a significant effect on the final result. Though these faint effects are not visually observable in the SDSS images, due to their low surface brightness, they do bias the parametrization algorithm towards more extended profiles. To avoid this artificial extension, we visually fine-tuned the minimum counts each pixel must have to be included in the parametrization algorithm. Also, given the high-density environment of the cluster, these parametrizations were occasionally affected by the presence of nearby galaxies. In these cases, global quantities, such as average ellipticity and position angle, was based on the observations of the inner region of the galaxies. Where possible, the photometric position angle of the galaxies was manually changed to match the observed kinematic position angle of the galaxy.

Furthermore, the presence of non-axisymmetric features, such as bars, can affect the results of the fit and force it away from the underlying stellar distribution. We reduce this effect by adapting the prescription described in \citet{Scott2013}. As per this, following an initial fit to the galaxies, we iteratively constrain the allowed minimum axial ratios of the Gaussians being fitted to the galaxies until the subsequent change in the mean absolute deviation of the fits grows by $>10\%$. This step is repeated while constraining the maximum allowed axial ratio. By doing so, the MGE model is pushed away from modelling these perturbations and instead models the underlying dominant light, and hence stellar distribution. This method significantly improves the dynamical models, providing improved predictions of the galaxy kinematics and better predictions of the galaxy inclination when compared to that inferred from the shape of the galaxy disk \citep{Scottetal2009_SAURON14}.

In this study, we use the light profile models generated from the MGE technique to measure the total magnitudes, effective radii (\re) and the major axis of the isophote containing half the light (\remaj) of the galaxies in our samples. The effective radii of the galaxies, \re, is determined as the radius containing half the light of the model when the model Gaussians have been circularized. \citetalias{atlas3d15} found that the effective radii, derived using the same photometric data (SDSS $r$-band photometry) and technique, underestimated the quantity derived by 2MASS, when 2MASS was calibrated to the RC3 catalog \citep{deVaucouleursetal1991}. To correct this inconsistency, the authors recalibrated their derived sizes by a factor of 1.35. In this study, we adopt the same correction for measured sizes for our galaxies.

\subsection{Models for Gravitational Potential}
\label{mass_models}

In order to model the stellar kinematics of galaxies, the underlying gravitational potential traced by the stars is required. However, much about the distribution of matter in galaxies is unknown, and it is not possible to confidently model ``realistic'' gravitational potential for galaxies. Hence to ensure robust interpretation of the results, in the following sub-sections we present a diverse set of models for this gravitational potential. 

It is important to note here that though the stars and dark matter dominate the gravitational potential of galaxies, this is not true in their innermost regions. In the innermost parsecs of galaxies, the gravitational potential of galaxies are dominated by their Supermassive Black Hole (SMBH). Here we model them using the M$_{\rm BH}-\sigma_{e}$ relation \citep{FerrareseMerritt2000,Gebhardtetal2000} with the parameters given by \citet[eq.~(7)]{KormendyHo2013}. However, to further ensure that these SMBHs do not affect the results presented in this study, we have masked the spaxels within the inner 1$''$ of the galaxies, which at its assumed distance of 100 Mpc \citep{Carteretal2008} effectively masks the inner \textasciitilde{0.8} kpc. For the two BCGs of the cluster, for whom we have high-resolution photometric data and for which R$_{BH}$ is resolved, we do not apply this mask and allow the mass of the SMBH to be a free parameter in the models. 

\subsubsection{Self-Consistent Model}
\label{gp_sc_sec}

The first model that we use for dynamical modelling is the self-consistent model, which assumes that the total mass distribution of a galaxy follows the distribution of its light. Such a dynamical model of the galaxies can be described by just three free parameters: (i) the inclination of the galaxy, (ii) the velocity anisotropy, defined as $\beta_z (r) = 1 - (\frac{\sigma_{z} (r)}{\sigma_{R} (r)})^{2}$ where $\sigma_z$ and $\sigma_R$ are the velocity dispersion along the axis of symmetry and radial axis of the galaxy, and (iii) the Mass-to-Light ratio, \mljam. Optimizing the first two parameters predicts the shape of the stellar kinematics, while the (M/L)$_{\rm JAM}$ scales the stellar kinematics to match that observed.

The deprojection of the observed surface brightness into a corresponding three-dimensional density distribution, needed to model and predict kinematics, is a mathematically degenerate problem \citep{Rybicki1987_Deprojection_degeneracy}. This problem is generally solved by ad hoc assumptions on the intrinsic shape of the galaxies. However, the MGE parametrization, for a given inclination, provides a unique solution for this deprojection. It does so by enforcing ellipsoidal/spheroidal equidensity and is known to provide luminosity densities that resemble realistic surface brightness when viewed at any angle. Concurrently, the MGE parametrization places limits on the possible values of inclination of galaxies, to permit realistic density distributions. For an axisymmetric galaxy, one can constrain the possible range of inclinations by requiring that the minimum axial ratio of every Gaussian be larger than a small value, like 0.05. This is done using eq.~(10) of \citet{cappellari2002}:

\begin{equation}
q^{2} = \frac{q'^{2} - \cos^{2}i}{\sin^{2}i}
\label{mge_inc}
\end{equation}

\noindent here, $q'$ is the minimum axial ratio of the Gaussians used to fit the galaxy photometry, $i$ is its inclination ($i=90^{\circ}$ is edge-on), and $q$ is the intrinsic axial ratio of the galaxy.

The velocity anisotropy describes the shape of the velocity ellipsoid. Though in real galaxies, the velocity anisotropy is known to vary radially across a galaxy in this study, we fix the $\beta_z$ of a galaxy to be constant. This assumption was tested by \citet{jam} and \citet{lablanche2012}, where the authors concluded that this assumption had a minor effect on the derived (M/L) and inclinations of the galaxies within the limitations of the JAM models. More recently, \citet{Lietal2016} assessed the ability of the JAM models to accurately model the kinematics of 1,413 realistic galaxies in the ILLUSTRIS simulation \citep{Vogelsbergeretal2014}. They conclude that the models recovered the ``total'' density profiles of the galaxies with good accuracy and negligible bias. More recently, an extensive study of 54 galaxies using the Asymmetric Drift Correction method, Jeans models (JAM), and Schwarzschild models by \citet{Leungetal2018} found that Jeans models and Schwarzchild's orbit superposition models reproduced the dynamical mass of galaxies in 1\re within 20\% scatter. In this study, we limit the values of $\beta_z$ for our models to between 0 and 1 as previous studies, using the more general Schwarzschild's orbit superposition models \citep{Schwarzschild1979}, demonstrate that the $\beta_z$ is well within this range \citep{sauron10,Thomasetal2009}. 

As mentioned earlier, the inclination and velocity anisotropy of the galaxies constrain the shape of the stellar kinematics, which can then be matched to the observed stellar kinematics of the galaxies to calculate the dynamical (M/L) of the galaxies. However, for the full sample, no information on the shape of the stellar kinematics is available. Hence for these galaxies, we have assigned default values to their inclination and velocity anisotropy. The inclination of these galaxies has been set to $60^{\circ}$, which is the average inclination for a set of randomly oriented galaxies. If an inclination of 60$^{\circ}$ is inconsistent with Eq.~\ref{mge_inc}, then the lowest permitted inclination for the galaxies is used as its inclination. The velocity anisotropy for these galaxies is set to 0.2. This value is driven by empirical measurement and detailed Schwarzchild modelling of the velocity anisotropy of 48 local elliptical and S0 galaxies \citep{sauron10}.

\subsubsection{Abundance-Matched Model}
\label{mass_model_am}

For this model, the total mass profile is described as the sum of the stellar and dark matter distribution. The shape of the stellar distribution is defined by the surface brightness of the galaxies, while the dark matter is a spherical NFW profile \citep{NFW_DarkHalo_1996}. For a given stellar M/L, i.e. \mlam, we can determine the stellar density profile and total stellar mass of the galaxy using the MGE parameterization of its photometry. The relation between the stellar-to-halo mass relation (SHMR) of \citet{MosterNaabWhite2013} and $c_{200}-M_{200}$ relation by \citet{KlypinTrujillo-GomezPrimack2011} then constrain the parameters of the dark matter halo and give the density profile of the halo. Hence, the \mlam of a galaxy constrains the total density profile, and the gravitational potential, of the galaxy. Since both relations used to constrain the dark matter profile of the galaxies are based on cosmological simulations, this model is heavily reliant on the theoretical assumptions of abundance matching.

As with the previous model, we solve for the best-fitting dynamical models of the SAURON sub-set galaxies by optimizing for the three parameters of the galaxies: (i) the inclination of the galaxy, (ii) the velocity anisotropy ($\beta_{z}$), and (iii) Stellar Mass-to-Light ratio (\mlam). The solution for the full sample is also derived similarly to the previous model, i.e., by fixing the values of the inclination and $\beta_z$ and optimizing for \mlam.

\subsubsection{Power-Law Model}
\label{mass_model_pl}

The third dynamical model is motivated by the empirical results of lensing studies \citep[see review by][]{Treu2010} and dynamical modelling \citep{Cappellarietal2015}. These results suggest that the total density profile of galaxies, within the region that this study samples, is well reproduced by a single nearly-isothermal profile, $\rho_{tot} \propto r^{-\gamma}$. We model this gravitational potential using a generalized NFW profile, i.e.:

\begin{equation}
\rho_{tot}(r) = \rho_{s} \left(\frac{r}{r_{s}}\right)^{\gamma} \left(\frac{1}{2} + \frac{1}{2}\frac{r}{r_{s}}\right)^{-\gamma - 3},
\label{PL_eq}
\end{equation}

\noindent where $\rho_{s}$ is the density of the galaxy profile at scale radius, $r_{\rm s}$. The break in the profile takes place at large radii, well outside the region sampled in the study, and hence, this profile is indistinguishable from that of a pure power-law, and its shape is affected only by $\gamma$. Also, the kinematics analysed in this study cannot reliably constrain the $r_{\rm s}$ of the profile as the quantity has minimal effect on the kinematics in the region studied here. Hence, we fix its value to 20 kpc, the median value for all ETGs in \atl ~as per their \textit{E} dynamical model. Therefore, the predicted stellar kinematics of this model is defined by only four free parameters: (i) galaxy inclination, (ii) velocity anisotropy ($\beta_{z}$), (iii) $\rho_s$, and (iv) slope of the profile ($\gamma$).

This assumption on the total density profile is only used to model the two-dimensional stellar kinematics of the SAURON sample as the one-dimensional kinematic data of the full sample cannot constrain the $\rho_s$ and $\gamma$ parameters of the dynamical model.

\subsection{Dynamical modelling}

With an MGE parametrization for the observed surface brightness for the galaxies in the cluster, and a model for their gravitational potential, we can now predict their second moment of velocity, $\overline{v_{los}^2}$, using the Jeans Anisotropic Models (\textsc{JAM})\textsuperscript{\ref{noteMCsoftware}} method \citep{jam}. For this study, we used version 5.0.9 of the code on Python 2.7. This code solves the Jeans equation for axisymmetric galaxies while assuming a velocity ellipsoid flattened along the axis of symmetry as observed in local galaxies \citep{sauron10, Thomasetal2009}. By comparing the predicted $\overline{v_{los}^2}$ with the observable $V_{\rm rms}=\sqrt{V^{2} + \sigma^{2}}$, where $V$ and $\sigma$ are the velocity and velocity dispersion, we can determine the best-fitting dynamical model that describes the observed galaxy. 

\subsubsection{SAURON Subset}
\label{sauron_dyn_models}

For every gravitational potential model discussed in Section~\ref{mass_models}, we optimize the free parameters of the model to reproduce the observed $V_{\rm rms}$ fields of the galaxies in the SAURON subset. However, for most galaxies, the observed kinematics maps appear noisy, which is inconsistent with the observed smooth galaxy images. Hence to reduce the effect of outliers values in the observed kinematic fields, we symmetrize these fields. We adopt the symmetrization technique used by  \citet{Cappellarietal2015}. Briefly, this technique duplicates every observed $V_{\rm rms}$ at position ($x_{i}, y_{i}$) to the positions ($-x_{i}, y_{i}$), ($x_{i}, -y_{i}$) and ($-x_{i}, -y_{i}$). This newly generated set of data points is then smoothed using the two-dimensional LOESS technique of \citet{cleveland1979} as implemented by \citetalias{atlas3d15}. In doing so, the technique exaggerates symmetric statistical features of the stellar kinematics while creating smooth kinematic fields, which are to be expected given the smooth photometry of the galaxies. It is to these fields that we fit to generate dynamical models of the galaxies. To further clean our kinematic fields during the modelling process, we create initial dynamical models and clip data points with absolute residuals greater than 2.6 times the bi-weighted standard deviation, i.e., we clip the spaxels with the largest 1\% of the residuals. This process is repeated until the process converges on a consistent set of data points, ensuring the models are not affected by erratic data. In the case of our two BCGs; due to the high S/N of the observed IFU datacubes, we do not symmetrize the kinematic fields for the dynamical modelling process. 

To determine the best fit models for our SAURON subset we optimize for the free parameters of the dynamical models presented in Sections.~\ref{gp_sc_sec}--\ref{mass_model_pl} using a Bayesian inference technique. In particular, we compute the posterior distribution of the free parameters of each model using the \textsc{emcee} \footnote{http://dan.iel.fm/emcee/current/} \citep{ForemanMackeyetal2013_emcee} implementation of the affine-invariant Markov-Chain Monte Carlo (MCMC) proposed by \citet{goodman2010ensemble}. The prior distribution of the free parameters is a constant within given reasonable limits and falls sharply beyond them. Assuming gaussian errors, we model the probability of a model for a set of parameters as:

\begin{equation}
P(data|model) \propto {\rm exp}\left(\frac{-\chi ^2}{2}\right),
\label{eq:}
\end{equation}

\noindent where the $\chi ^2$ is defined as 

\begin{equation}
\chi ^2 = \sum_{j} \left( \frac{\left\langle v_{los}^{2} \right\rangle _{j}^{1/2} - V_{rms,j}}{\Delta V_{rms,j}} \right) ^2, 
\label{eq:}
\end{equation}

\noindent here $\left\langle v_{los}^{2} \right\rangle _{j}^{1/2}$ is the predicted second moment of velocity for the dynamical model being evaluated, while $V_{rms,j}$ and $\Delta V_{rms,j}$ are the observed stellar kinematic and its error, in the $j^{th}$ bin of the galaxy's two-dimensional kinematic map. The best fit values of the free parameters, and their associated uncertainties, are derived from their respective posterior distributions.

In order to test the quality of the dynamical models, we visually compared the observed kinematics with that predicted by the best-fit models. This comparison was done between the observed V$_{rms}$ maps, post symmetrization, and the $\left\langle v_{los}^{2} \right\rangle ^{1/2}$ for the best fit dynamical model. Based on these plots we visually classify the data quality into three groups using similar classification as \citetalias{atlas3d15}: galaxies with good dynamical models that reproduce the observed kinematic structures (\textit{Qual} = 2 in Appendix.~\ref{Table_sauron}), galaxies acceptable dynamical models that reproduce some of the observed kinematic structures (\textit{Qual} = 1), and galaxies for whom reliable kinematics could not be derived (\textit{Qual} = 0). For the visual classification, we have focused solely on the reproduction of kinematic features that are sufficiently far from the center that they are not affected by an inaccurate estimate of the SMBH. 

\begin{figure*}
	\centering
	\includegraphics[width=1.9\columnwidth]{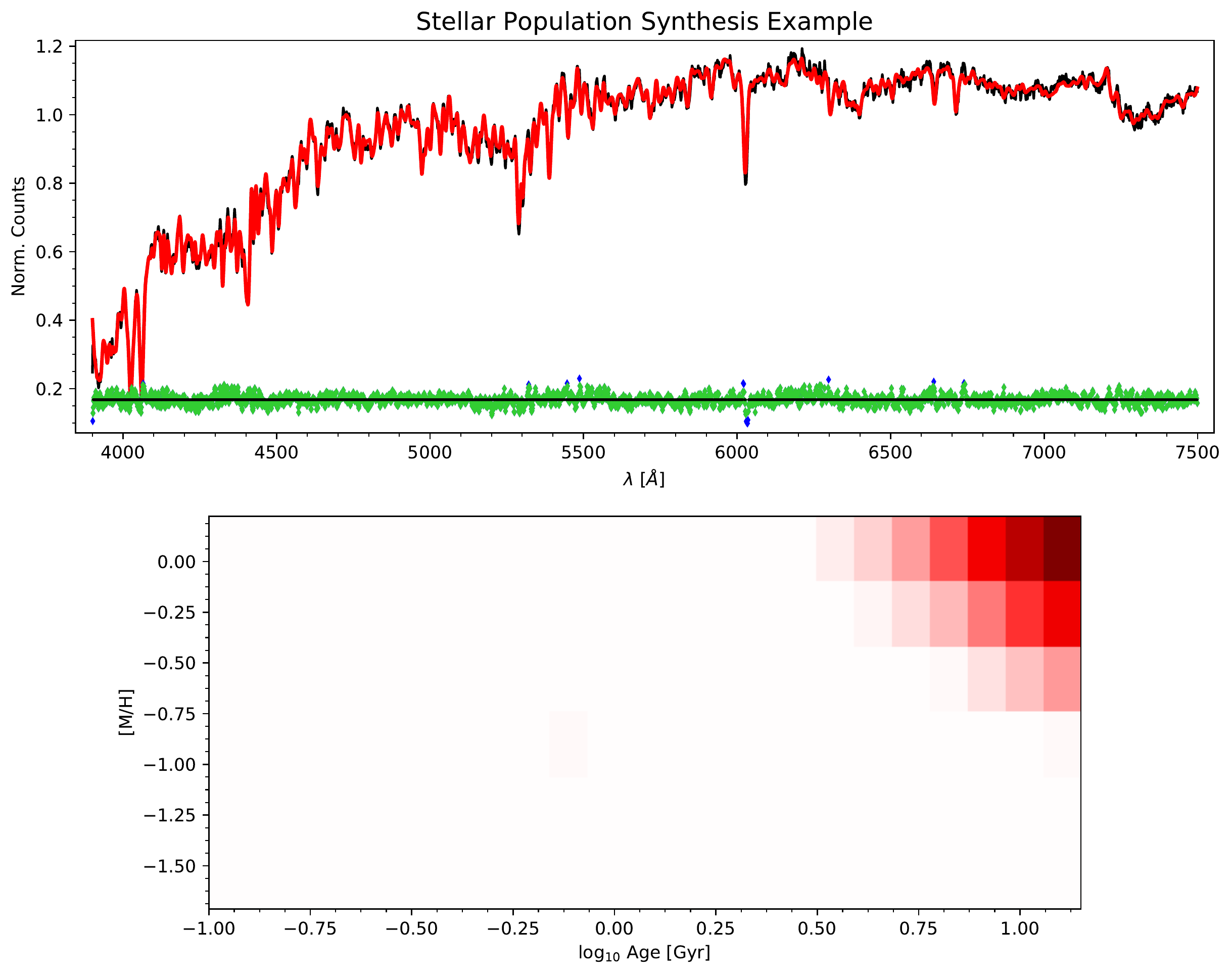}
	\caption[SPS example]{An example of the typical quality and result of the full spectrum fitting of the SDSS spectrum with the MILES stellar population models for IC3943. In the top panel, we present the observed galaxy spectrum in a solid black line with the best-fit spectrum overlayed in red. The green points in the panel present the residuals for the fit, offset to ease examination, with the blue points representing the same for pixels clipped during the fitting process. The panel demonstrates that the full spectrum fitting reproduces the observed galaxy spectrum very well. The lower panel illustrates the distribution of weight in the best-fit solution for the different stellar population models, with darker shade representing higher relative weight for the SSP of a given age and metallicity. Together the plots demonstrate that the regularization of the weight distribution allows the code to reproduce the observed galaxy spectrum while simultaneously ensuring a smooth star formation history for the galaxies.}
	\label{fig:SPS_example}
\end{figure*}

\subsubsection{BCGs: NGC4874 \& NGC4889}

For the two BCGs, NGC 4874 and NGC 4889, additional steps were taken to ensure high-quality dynamical models can be created. For these dynamical models, we replace the MGE results taken from the SDSS photometric data with that from the higher-resolution \textit{HST} data. These new models permit us to generate higher resolution dynamical models of the galaxies in the region covered by the IFU. 

Moreover, these galaxies are expected to be weakly triaxial, and hence modelling them with the assumption of axisymmetry is not realistic. Therefore for these galaxies, we created models using the code \textsc{jam\_sph\_rms}, which is described in Section~3.2 of \citet{jam}. In these models the surface brightness and mass profile of the galaxies are circularized and, since these galaxies have minimal rotation, we assume that the velocity dispersion of the galaxies is a close approximation of the actual $V_{\rm rms}$ of the galaxies as slow rotators are close to spherical systems (e.g.: \citetalias{Cappellari2016_review}).

\subsubsection{Full Sample}

In the case of the full sample, we generate dynamical models using the self-consistent and abundance matched models to determine the \mljam ~and \mlam ~for the galaxies. These models are generated using the \textsc{JAM} algorithm. Using the observed light profile, and assumed mass model, velocity anisotropy and inclination for the galaxies, as presented in Section.~\ref{surf_brightness_parametrization}-\ref{mass_model_am}, the algorithm can generate an expected V$_{rms}$ that a spaxel of radius equivalent to that of the SDSS fibre would observe. The expected V$_{rms}$ can then be scaled to match the measured velocity dispersion of the galaxies by scaling the \mljam ~for the self-consistent models, and \mlam ~for the abundance matched models. For this latter stage, we use the Levenberg-Marquardt method which solves non-linear least-square problems \citep{levenbergmarquardt1,levenbergmarquardt2}, implemented using \textsc{LMFIT} code in Python \citep{lmfit_python}. 

The use of these simple dynamical models over the virial relation is motivated by the observed sensitivity of the latter on the technique used to measure \re ~of galaxies. This sensitivity is demonstrated in section.~4.4 of \citetalias{atlas3d15} wherein the authors demonstrate that dynamical masses determined using simple virial estimators vary systematically compared to that measured using detailed dynamical models. Through our technique, we can account for the complex light profiles observed in real galaxies, more so than simple S\'{e}rsic profile, and can include physically motivated velocity anisotropy and intrinsic flattening. As a consistency test, we compared the \mljam ~measured by the technique against the same measured by the detailed dynamical modelling of the observed SAURON data for 15 galaxies with $Qual = 2$ (i.e. the observed kinematic structures of the galaxies are well reproduced by the models). The linear relation between the log of the quantities has a slope of 1.09$\pm$0.17, and a scatter of 0.07dex. 

\subsection{Stellar Population modelling}

The modelling of the stellar populations of these galaxies was done via the full-spectrum fitting technique using the \textsc{pPXF} code and single stellar population models. However, the solution for the star formation history of galaxies is degenerate, and hence any modelling of the underlying stellar populations of galaxies is dependent on the fitting technique. In this study we adopted to regularize the weights (see \citealt{ppxf_2}) associated with population models along the template grid until the $\chi^{2}$ of the galaxy spectrum fit increases with respect to an un-regularized fit by $\sqrt{2 \times N_{DoF}}$ \citep[section 19.5]{NumericalRecipes}. This technique guides the full spectrum fitting algorithm away from unrealistic formation histories (with random isolated formation events) to a more realistic formation history wherein stars form continuously and gradually in a galaxy. Simultaneously this technique allows for the inclusion of a second burst in star formation if the data requires one. For a detailed description of the technique, we refer the reader to section~2.5 of \citet{ATLAS3D30}.

For these galaxies, we model the stellar populations using the MILES stellar population models \citep{Vazdekisetal2015_MILESModels} which are based on the MILES empirical stellar spectrum library \citep{Miles_stellar,Falcon-Barrosoetal2011}. The template grid covers an age range of 0.08~--~14.1 Gyrs in 22 steps sampled logarithmically, and six metallicities; -1.71, -1.31, -0.70, -0.39, 0.00, \& +0.22. Despite the availability of stellar population models up to ages of 17.78 Gyrs for this study we limit our templates to 14.1 Gyrs, the approximate age of the universe since it is unrealistic for galaxies to host older stellar populations. The stellar population models span a wavelength of 3,540~--~7,410\AA. We model the stellar population of the galaxies using their SDSS spectrum, however, due to limitations in the wavelength range of the models, only 3,900~--~7,500~\AA ~of the SDSS spectra are used. These models assume a Salpeter IMF \citep{salpeter1955} and use 0.1 M$_{\odot}$ and 100 M$_{\odot}$ lower and upper stellar mass cut-offs respectively. 

Since the MILES population models are based on the MILES empirical stellar spectra library, the models have a resolution limit of 2.54\AA~as well. However, the instrumental resolution of the SDSS spectra varies significantly across their wavelength coverage, with significant portions having a higher resolution than the MILES models. To account for this variation during the stellar population synthesis, we broadened the observed galaxy spectrum as a function of wavelength, using the \textsc{gaussian\_filter1d} module provided with \textsc{pPXF}, to match the resolution of the MILES models across the observed wavelength range. To test the effect of this approach, we compared the results against the results of stellar population modelling without the correction for the varying resolution and found no significant differences.

In Fig.~\ref{fig:SPS_example} we present an example of results of the full-spectrum fit to the observed galaxy spectra and the weight distribution of the templates used along a grid of model age and metallicity. These plots demonstrate that our stellar population modelling technique can reproduce the observed galaxy spectrum with relatively small residuals while ensuring a smooth star formation history for the galaxies.

Based on the resulting weights of the regularized spectral fitting, we derive the stellar \mlsalp and age of the stellar populations using:

\begin{equation}
(M_{\ast}/L)_{\rm Sal}= \frac{\sum_{j}w_{j}M_{\star, j}^{\rm no gas}}{\sum_{j}w_{j}L_{{\rm B},j}},
\label{MLeq}
\end{equation}

\begin{equation}
\langle\log({\rm Age})\rangle = \frac{\sum_{j}w_{j}\log({\rm Age}_{j})}{\sum_{j} w_{j}},
\label{age_eq}
\end{equation}

\noindent where $w_{j}$ is the weight attached to the $j^{th}$ stellar population model by the regularized mass-weighted fit to the galaxy spectrum, $M_{\star, j}^{\rm no gas}$ is the mass of the model that is in stars and stellar remnants, $\log({\rm Age}_{j})$ is the log of the age of the model and $L_{{\rm r},j}$ is the \textit{r}-band luminosity of the template. 

\begin{figure}
	\centering
	\includegraphics[width=0.9\columnwidth]{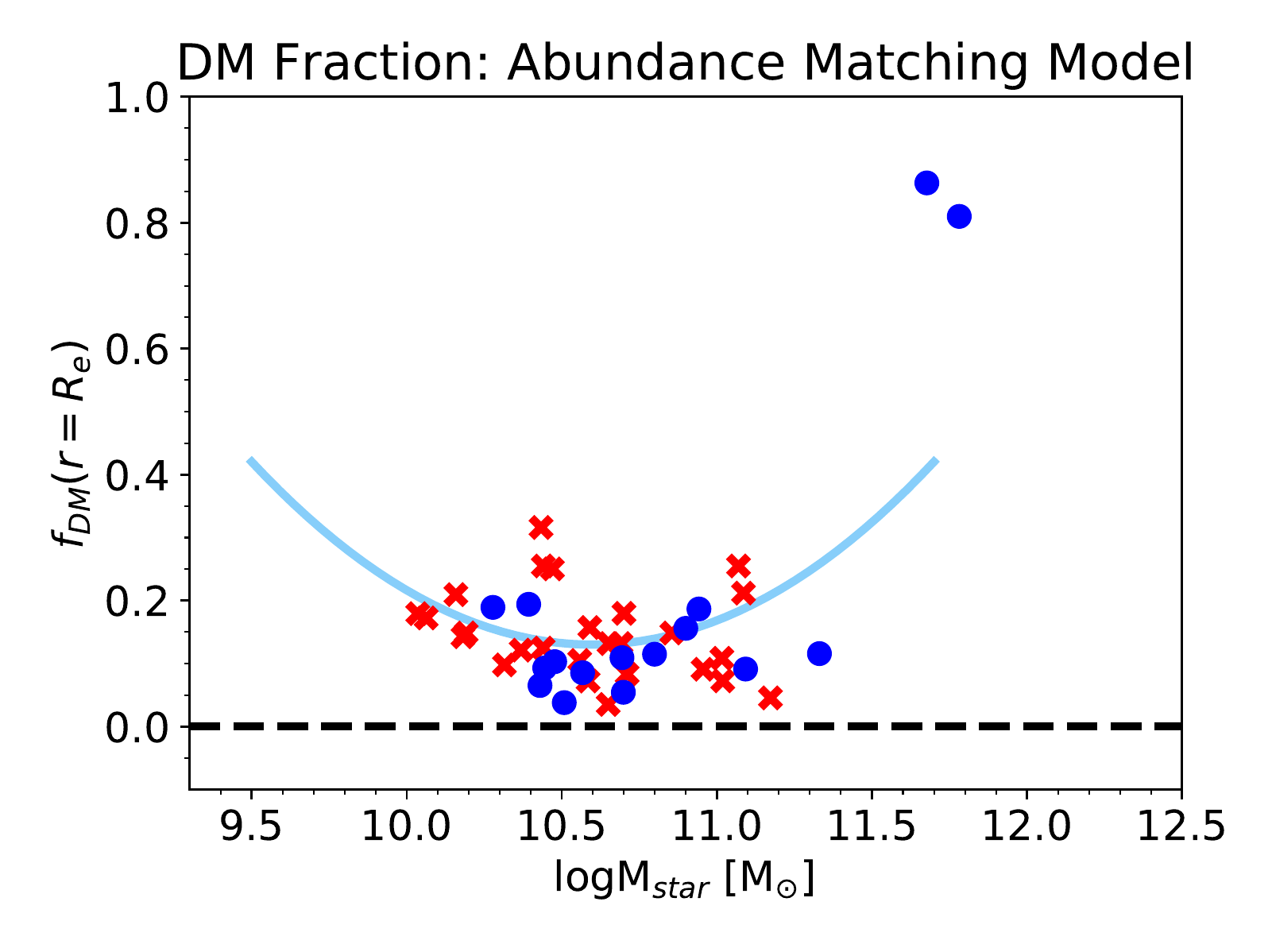}
	\includegraphics[width=0.9\columnwidth]{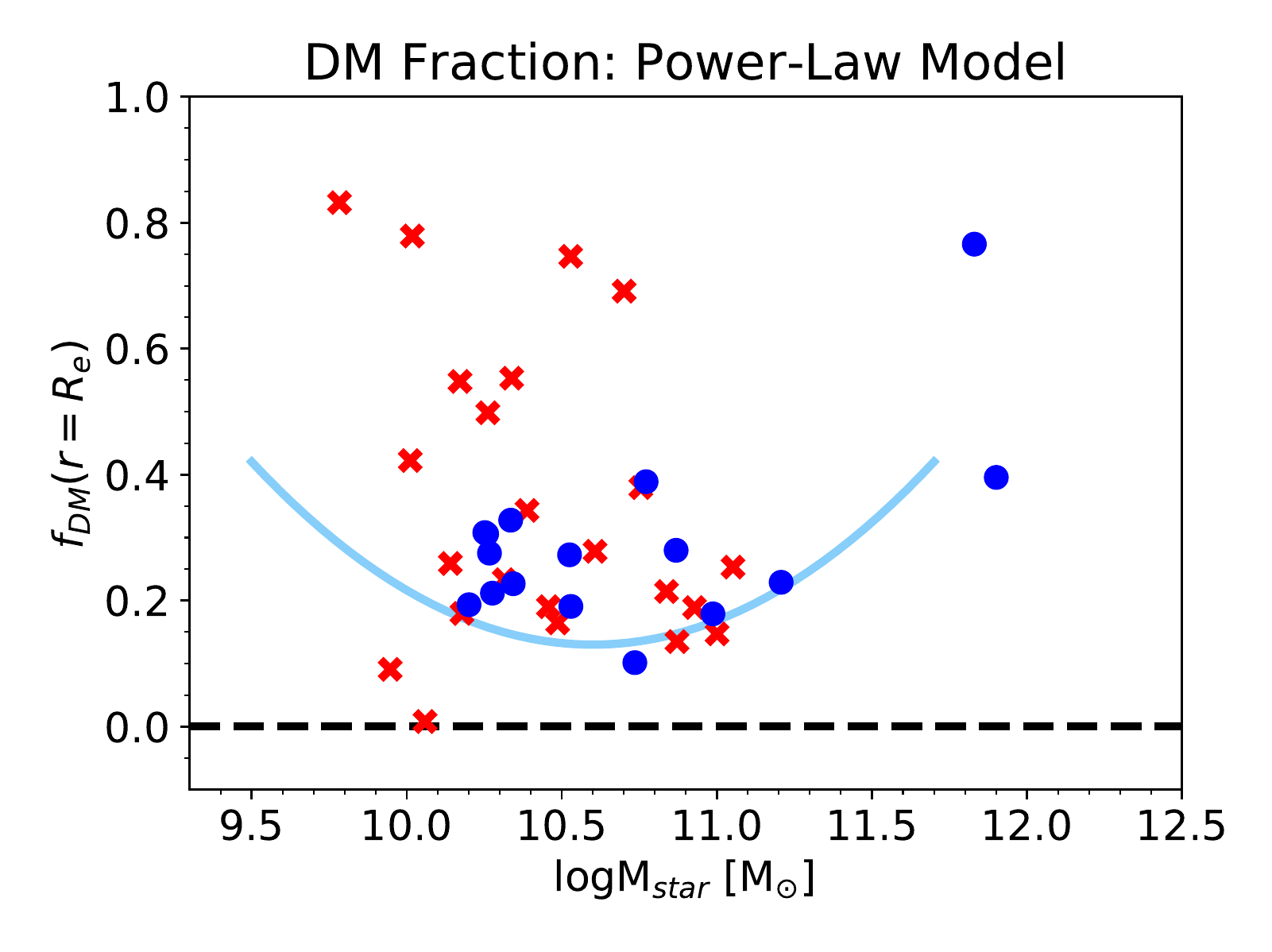}
	\caption[Dark Matter Fraction of the Coma Cluster Galaxies: SAURON Subset]{The dark matter fractions within a sphere of radius 1\re~for the SAURON subset using the IFU data, calculated from the best fit dynamical models using the Abundance Matching model (top) and the Power Law model (bottom). The blue circles represent the galaxies with \textit{Qual} = 2 and the two BCGs of the cluster, while the red crosses represent the rest. The solid blue line represents the curve fit to the dark matter fraction derived for the \atl ~~galaxies using a model similar to that of the Abundance Matching model in this study. A comparison between the dark matter fractions derived from the two distinct models, robustly indicates that the dark matter fraction of these galaxies are low. However, the plot presents a discrepancy between the dark matter fractions derived for NGC 4889 using the two models. Upon investigation, we find that we are unable to reproduce the observed spatially resolved stellar kinematics of the galaxies using the abundance matching model and hence highlighting a potential discrepancy between theory and observations.}
	\label{dm_frac}
\end{figure}

\begin{figure}
	\centering
	\includegraphics[width=0.9\columnwidth]{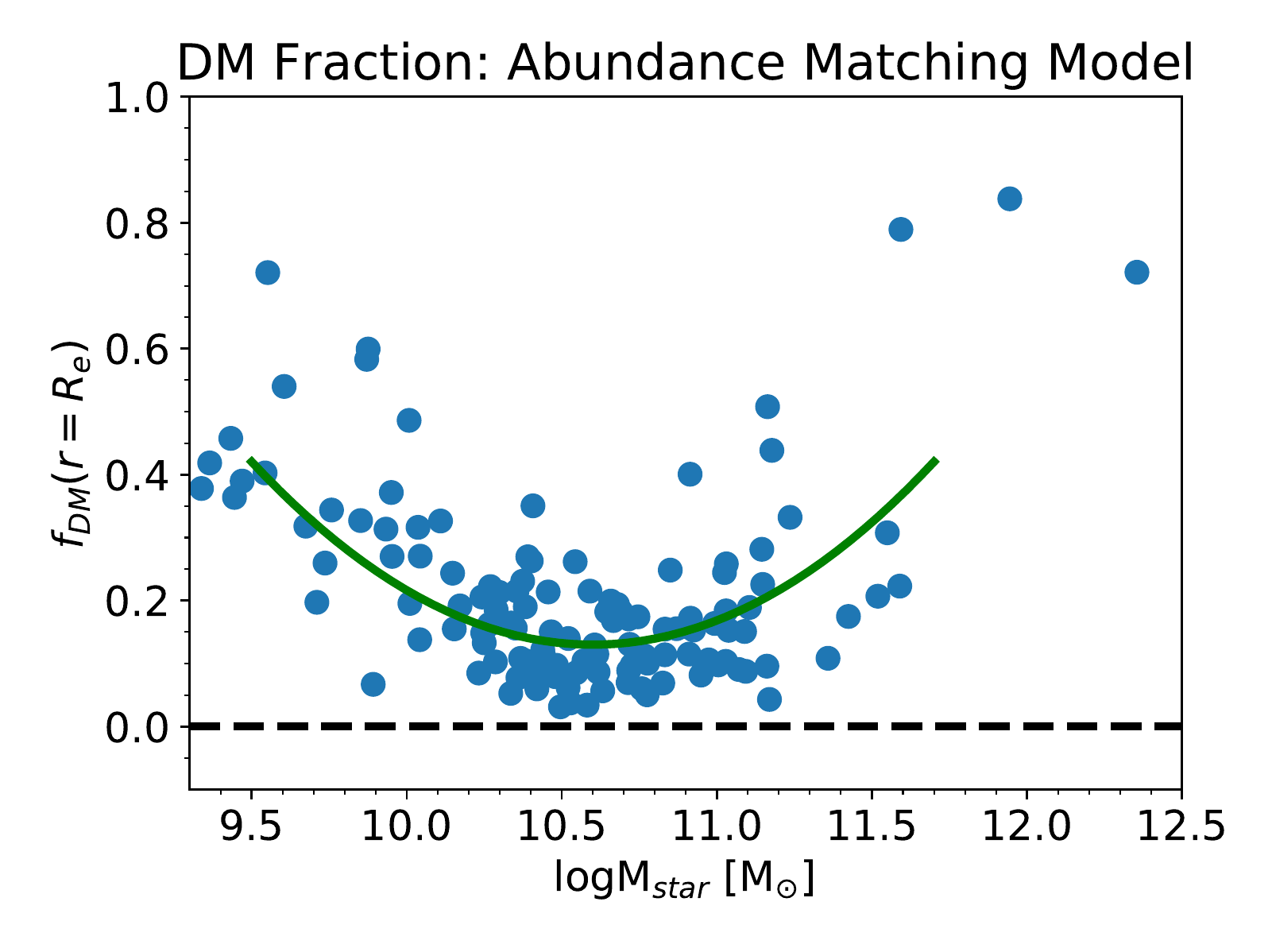}
	\caption[Dark Matter Fraction of the Coma Cluster Galaxies: Full sample]{The dark matter fraction within a sphere of radius 1\re ~as derived from the simple dynamical models using the one-dimensional SDSS spectra for the Full sample. The plot includes all 148 galaxies of the full sample, including the galaxies in the SAURON subset. The solid line in the plot represents the trend seen in the sample by \atl ~using a similar model for the total mass profile of the galaxies.}
	\label{dm_frac_sdss}
\end{figure}

\section{Results}

Using the above analysis, we have determined a number of physical quantities for our two galaxy samples. These quantities are presented in Appendix.~\ref{Table_sauron} and Appendix.~\ref{Table_full} for the SAURON subset and full sample, respectively. V$_{rms}$ field maps and MGE parametrization used for the dynamical modelling of the SAURON subset are available as supplementary material on the journal website.

\subsection{Dark Matter in Galaxies}
\label{subsec:dm}

In this study we have used two independent assumptions for dark matter in our SAURON subset of galaxies; AM (Abundance-Matched) model based on numerical simulations, and PL (Power-Law) model based on empirical observations. These model assumptions are sufficiently distinct to provide a sense of the expected systematic uncertainties in our dark matter determinations.

For the Abundance-Matched model, we use the best fit \mlam to define the dark matter halo of the galaxy, as described in Section.~\ref{mass_model_am}. The stellar mass distribution of the galaxies is given by the deprojection of their light profile multiplied by the \mlam. To quantify the dark matter content in galaxies, we measure their dark matter fractions which is the ratio of the mass of the dark matter halo to the total mass within a sphere of radius 1\re.

To calculate dark matter fraction for the PL model, one needs to the first disentangle the stellar and dark matter component from the total density distribution, which is fitted to observed data. To do this, we proceed as \citet{Mitzkusetal2017} and \citet{Pocietal2017} by defining the problem as
\begin{equation}
\rho_{tot}(r) = (M_*/L)_{dyn}^{PL}  \rho_*(r) + \rho_{\rm s}  \rho_{\rm DM},
\label{eq:total}
\end{equation}

\noindent where,
\begin{equation}
\rho_{\rm DM} = \left( \frac{r}{r_s} \right)^{-1} \left( \frac{1}{2} + \frac{1}{2} \frac{r}{r_s} \right) ^{-2} \qquad \gamma = -1,
\label{eq:nfw}
\end{equation}

\noindent here the best-fitting total density profile of the PL model, $\rho_{tot}(r)$, is defined as the linear sum of the stellar density profile and the dark matter halo of the galaxy. The stellar density profile is the observed light profile of the galaxy, deprojected using the best-fit galaxy inclination from the dynamical modelling, and scaled by the stellar M/L of the galaxy, i.e. \mlpl. For our dark matter halo, we assume an NFW profile \citep{NFW_DarkHalo_1996}, hence our dark matter halo is defined by the shape of the profile, Eq.~\ref{eq:nfw}, scaled by an amplitude $\rho_{\rm s}$. We disentangle the stellar and dark matter halo from our best-fit total density profile for the galaxy by optimizing for the two free parameters in Eq.~\ref{eq:total}, i.e. \mlpl and $\rho_{\rm s}$, that reproduces the best-fit total density profile. 

\begin{figure*}
	\centering
	\includegraphics[width=0.45\textwidth]{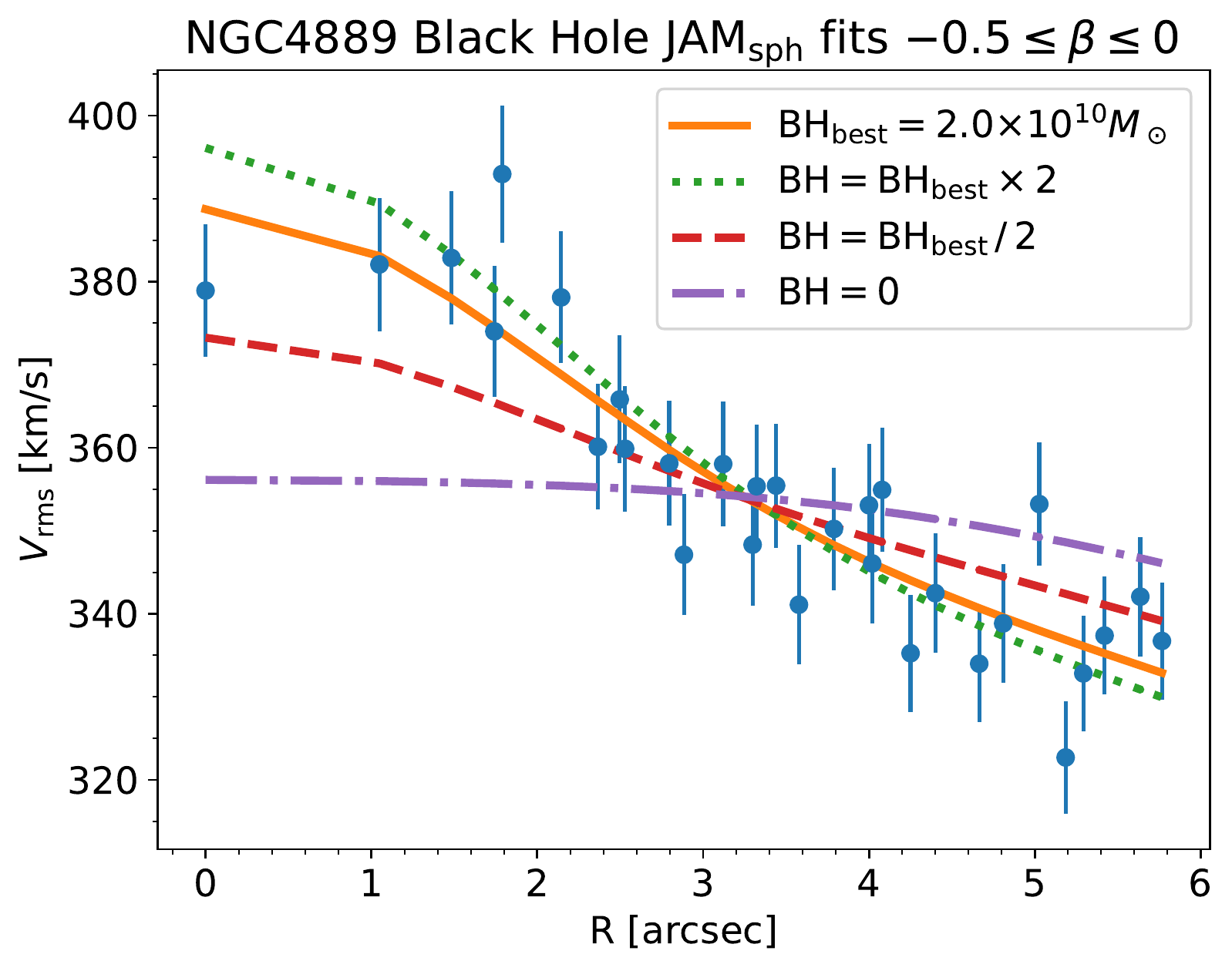}
	\includegraphics[width=0.45\textwidth]{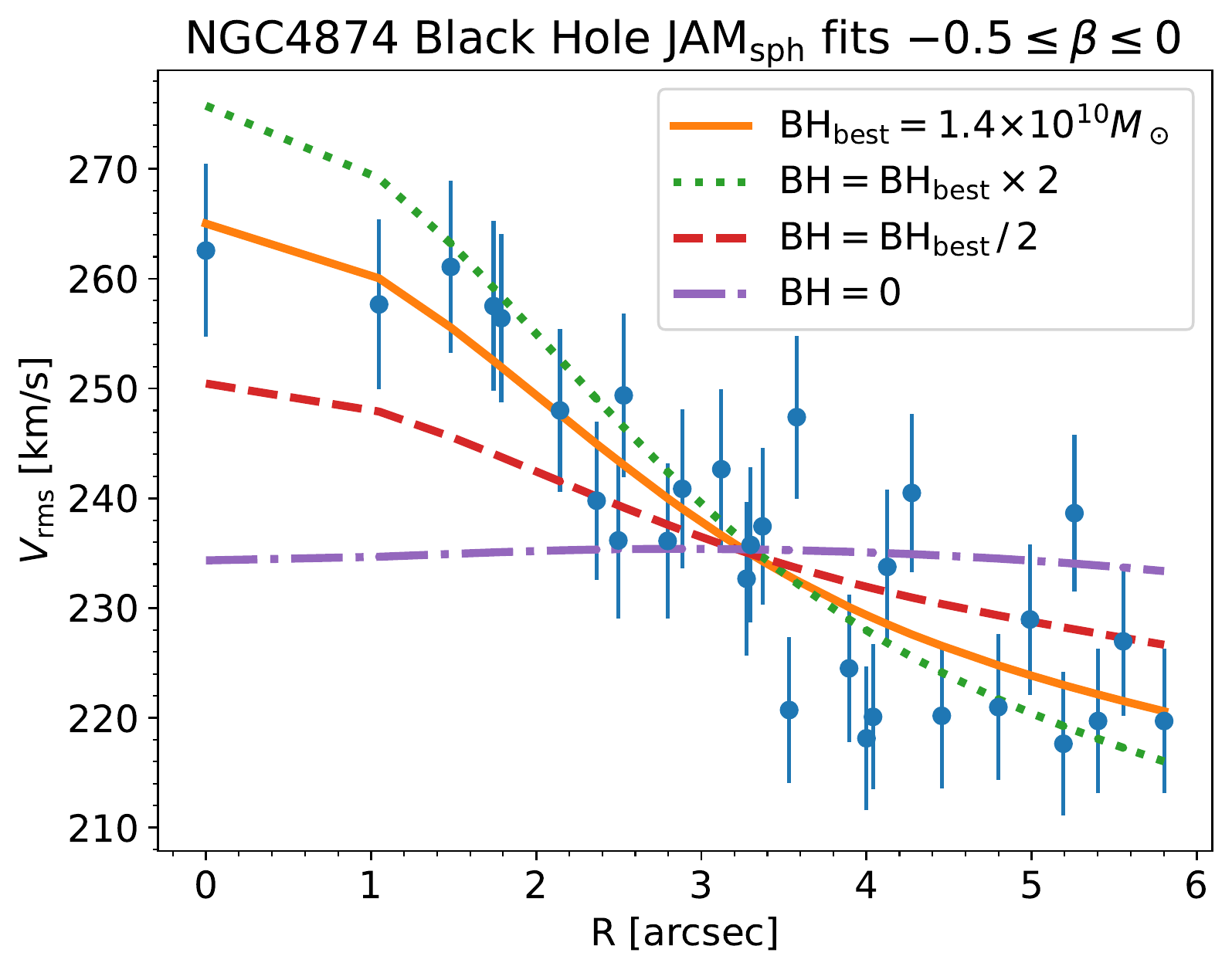}
	\caption[Dynamical Model for BCGs]{In the panels above we illustrate the radially averaged V$_{rms}$ ($= \sqrt{V^{\rm 2} + \sigma^{\rm 2}}$) for NGC4889 (left) and NGC4874 (right) in blue filled points. For illustration, the errors presented are normalized to give $\chi^2$/DOF$=$1 for the best fitting JAM model, shown in the solid orange line. The dotted green and dashed red lines demonstrate the best-fitting dynamical models for a BH of mass two times, and half of the best-fitting mass respectively. These models fit for the velocity anisotropy ($\beta$) and hence demonstrate the approximate uncertainty in the determined mass of the SMBHs of the BCGs. The dot-dashed violet line presents the best-fitting model with no BH at the centre of the galaxies and illustrates the necessity of the BHs in the models to reproduce the observed kinematics of the galaxies.}
	\label{BCGs_SMBH}
\end{figure*}

In Fig.~\ref{dm_frac}, the dark matter fractions for the SAURON subset are presented for both models. The plots robustly demonstrate that for most galaxies, the dark matter fraction within a \re ~is low. Quantifying this result for the Abundance Matched model, 90$^{\rm th}$ percentile of galaxies classified as having good models, have a dark matter fraction of \textasciitilde{50}\% and a median of 11.2\%. For the Power Law model, these quantities are 34.6\% and 25\% respectively. This result is consistent with the results found by \citet{Thomasetal2007} where the authors used long-slit data of 17 ETGs in the Coma cluster to create axisymmetric Schwarzchild models for the galaxies. These authors find that the dark matter fraction for their galaxies are mostly within 10-50\%, consistent with our determinations. These values are also consistent with those found by \citetalias{atlas3d15} and \citet{Pocietal2017} who used a diverse set of models for the dark matter halo, and total mass profile for the \atl ~galaxies respectively. Cosmological simulations have also predicted low dark matter fractions for the inner regions of galaxies \citep[e.g. ][]{Kobayashi2005, TaylorKobayashi2015}. However, our results are quantitatively inconsistent with these predictions. Further work is needed to investigate the origin of this inconsistency and is beyond the scope of this study. In Fig.~\ref{dm_frac_sdss}, we present the dark matter fractions determined using the abundance matched model for the full sample. The plot, consistent with results in Fig.~\ref{dm_frac}, demonstrates the low dark matter fraction for most galaxies in the full sample.

One source of uncertainty in these results is the effect of gradients in stellar populations within our observed early-type galaxies. Studies on the stellar population of early-type galaxies using IFU observations have demonstrated the presence of strong negative metallicity gradients \citep[e.g.:][]{SAURON14, GonzalezDelgadoetal2014, GonzalezDelgadoetal2015, Greeneetal2017MASSIVEstellarpop, Boardmanetal2017, Parikhetal2019, SAMIradialgradients}. These invariably lead to a non-constant M/L within the galaxies and hence could potentially have a significant impact on the derived stellar mass estimates \citep{Bernardietal2018} and on our measured dark matter fractions.

In order to study the effect of these gradients, we introduced a M/L gradient and recreated the dynamical models for the two samples of galaxies. The M/L gradient used in this test was based on the results of \citet{DominguezSanchezetal2019} where the authors studied gradients of various stellar population properties within elliptical and S0 galaxies of the MaNGA-DR15 survey \citep{MaNGADR15}. Within their sample, the authors find that galaxies with the highest central velocity dispersions and absolute \textit{r}-band magnitudes demonstrate the strongest but modest gradient, with M/L never varying more than a factor of two out to 0.8R$_{e}$. Since we intend to identify the effect stellar population gradients can have on the derived stellar M/L of the galaxies, we implement the steepest M/L gradient observed by the study to all galaxies in our two samples. For simplicity we ignore the variation of the M/L gradients observed to correlate with different galaxy properties and enforce a linearly varying radial M/L gradient.

The M/L gradient was incorporated into the density profile of the dynamical models by the rescaling the Gaussians of the MGE models describing the stellar mass profile of the galaxies (please refer to Section.~\ref{mass_models}) with the expected change in the M/L at the gaussian's dispersion. This gradient scaling was introduced for the entire MGE models which extend well beyond the region wherein the stellar kinematics are measured.

We find that a radial gradient in the stellar M/L within the field of view of our observations increases the average stellar M/L measurement by <10\% and average dark matter fraction within an R$_{e}$ by <0.02. These systematic offsets are well under the uncertainties on these quantities, and hence a systematic effect of M/L gradients can be ignored. However, the gradient could affect the observed scatter in the relations presented in this study. Unfortunately, accounting for this is beyond the scope of the available data and is hence ignored.

Next, we focus on the two BCGs in the cluster, NGC 4889 and NGC 4874, which are the two most massive galaxies in Fig.~\ref{dm_frac}. We observe that the dark matter fraction for NGC 4874 (the less massive of the two BCGs) is consistent for both sets of dynamical models, but for NGC 4889 we see a drastic difference between the models. A study of the observed $V_{\rm rms}$ and that predicted by dynamical models using the Abundance-Matched mass distribution demonstrates the cause as these models predicting an upturn in $V_{\rm rms}$ after $\lesssim$8" (3.9kpc) from the galaxy centre. This upturn is likely caused by an over-concentration of the predicted galaxy halo as at these radii the halo density dominates the mass density. Recently, there have been indications that the NFW profile may over-predict the halo concentration for massive halo density profiles \citep{Klypinetal2015,Shanetal2015}. In work by \citet{NewmanEtAl2013_luminous_DM}, the authors used strong and weak lensing and resolved stellar kinematics to disentangle the luminous and dark matter component in seven BCGs in relaxed clusters at z=0.2-0.3. These authors find that the inner dark matter profile for the galaxies is correlated with the distribution of stars in the BCG and that a cored-NFW profile, with a shallower profile than the standard NFW profile within \textasciitilde{\re}~ of the BCGs, provides an equally good model for their results. Using detailed dynamical models along with high-quality IFU data extending further than a \re~ could provide further clarity to this issue.

Given that the two BCGs, like most massive slow rotators, are weakly triaxial \citepalias[e.g.][]{Cappellari2016_review}, we used spherical rather than axisymmetric JAM models \citep{jam}. The fits were restricted to the smallest spatial region that still allowed us to break the degeneracy between M/L and BH mass \citep[as done, for the same reason in][]{Cappellaretal2010, Drehmeretal2015, Krajnovicetal2018, Thateretal2019}. In this way, our fits are more robust against possible gradients in anisotropy or stellar M/L within the very innermost regions. The SAURON kinematics was averaged in radial bins containing the same number of kinematic values, and we assumed constant errors in the fits. 

\begin{figure}
	\centering
	\includegraphics[width=\columnwidth]{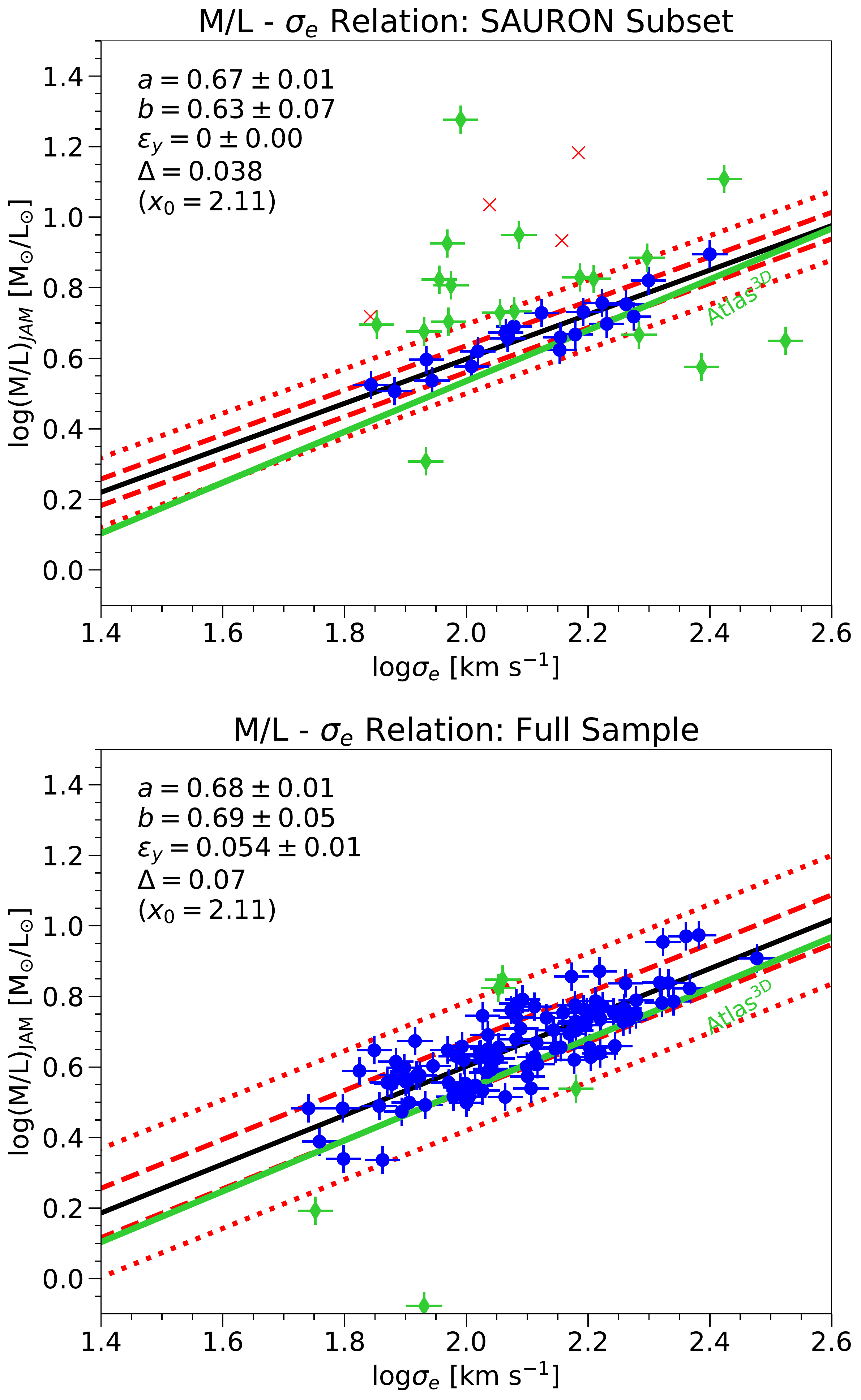}
	\caption[(M/L)-$\sigma_{e}$ Relation for Coma cluster]{The dynamical (M/L)~--~$\sigma_{e}$ relation for the SAURON subset using the IFU data (top) and the full sample using the one-dimensional SDSS data (bottom). The blue points in the top panel represent the galaxies with \textit{Qual} = 2, with the green triangles and red crosses depicting galaxies with acceptable (\textit{Qual} = 1) and bad (\textit{Qual} = 0) dynamical models respectively. In the top left portion of the panels we present the results of a line fit to the data using \textsc{LTS\_LINEFIT}; the slope (\textit{b}) and offset (\textit{a}) of the relation, the intrinsic scatter along the y-axis ($\epsilon_y$), the root-mean-square scatter of the relation ($\Delta$), and the pivot point used when fitting the linear equation  y = a + b(x - x$_0$). In both panels, a linear fitting was done to the blue points using a least trimmed square fitting technique, with the best-fitting quantities printed in the top left of each panel. The solid black line represents the best-fitting line to the points, while the dashed red and dotted red lines mark lines of 1$\sigma$ and 2.6$\sigma$ of the relationship, respectively. The solid green line in the panels represent the best-fit relationship found by the \atl ~survey.}
	\label{ml_vd_compare}
\end{figure}

\begin{figure}
	\centering
	\includegraphics[width=\columnwidth]{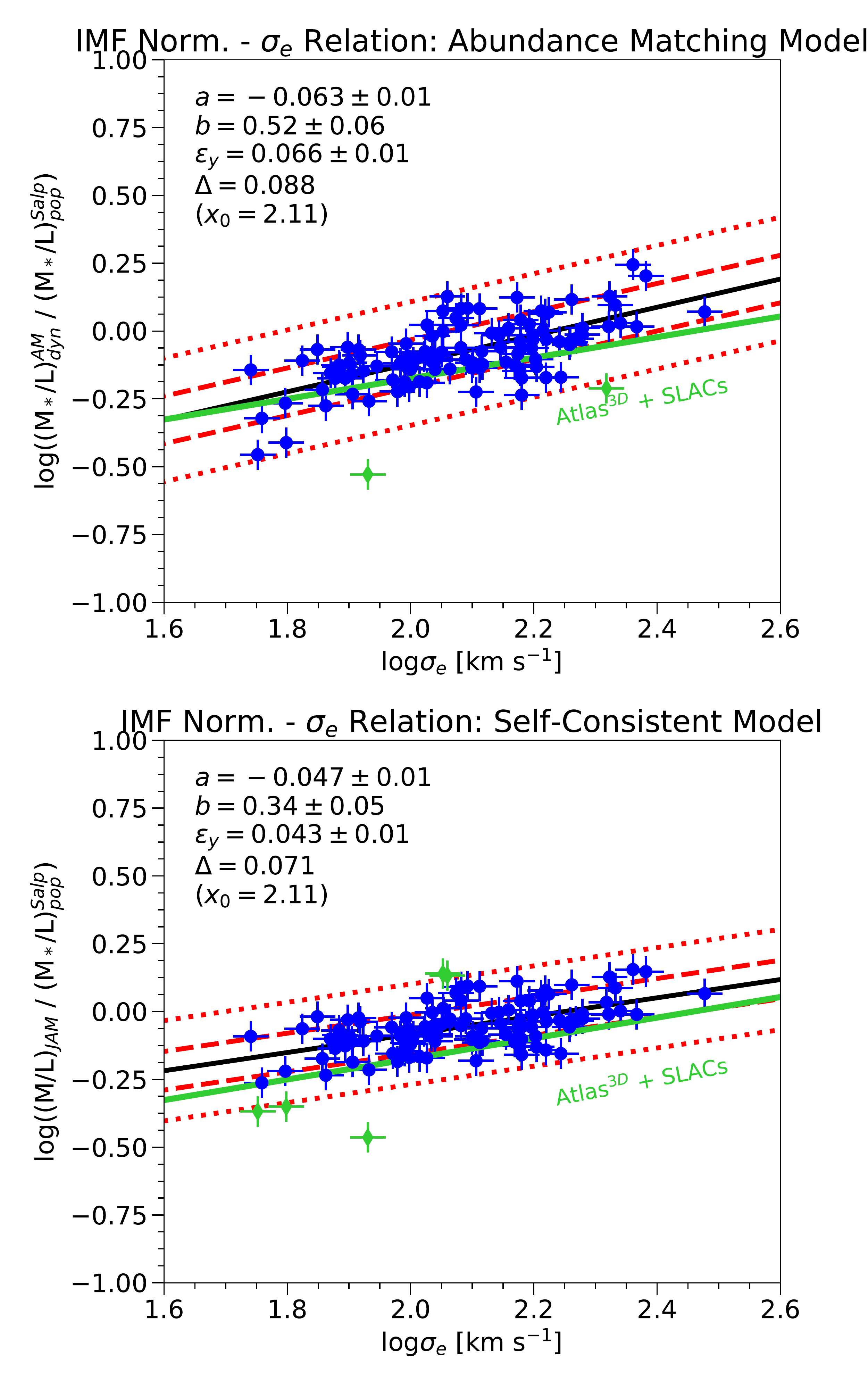}
	\caption[IMF Normalization-$\sigma_{e}$ Relation]{The IMF normalization~--~$\sigma_e$ relation for the galaxies in the Coma cluster. \textit{Top panel}: The IMF normalization is derived using the stellar (M/L)$_{dyn}^{AM}$ from the Abundance Matching model and the stellar (M/L)$_{star}^{Salp}$ calculated from the regularized mass-weighted fit to the galaxy spectra. The black line represents the best fit to the full sample, depicted in blue circles, and the dashed and the dotted red lines are the $1~\sigma$ and $2.6~\sigma$ lines respectively. The solid green line is the IMF normalization--$\sigma_e$ relation for the extended sample of the \atl ~+ SLACS galaxies. As in Fig.~\ref{ml_vd_compare}, the top left portion of the panels present the results of the linear fit to the data. As expected, due to the low relative uncertainty in distances, the galaxies in the Coma cluster illustrate a tighter IMF normalization -- $\sigma_e$ relation than previous studies. This relation appears to have a steep slope and a significant offset. \textit{Bottom panel}: This plot is identical to that in Fig.~\ref{imf_norm}, with the difference that here we derive the IMF normalization using the dynamical (M/L)$_{JAM}$ derived using the Self Consistent model. A comparison of the observed relation with that seen locally (\atl ~+ SLACS) demonstrates a consistent slope for the relationship, with a small difference in offset. This offset may be caused by the difference in SSP models used during stellar population synthesis.}
	\label{imf_norm}
\end{figure}

\begin{figure}
	\centering    
	\includegraphics[width=\columnwidth]{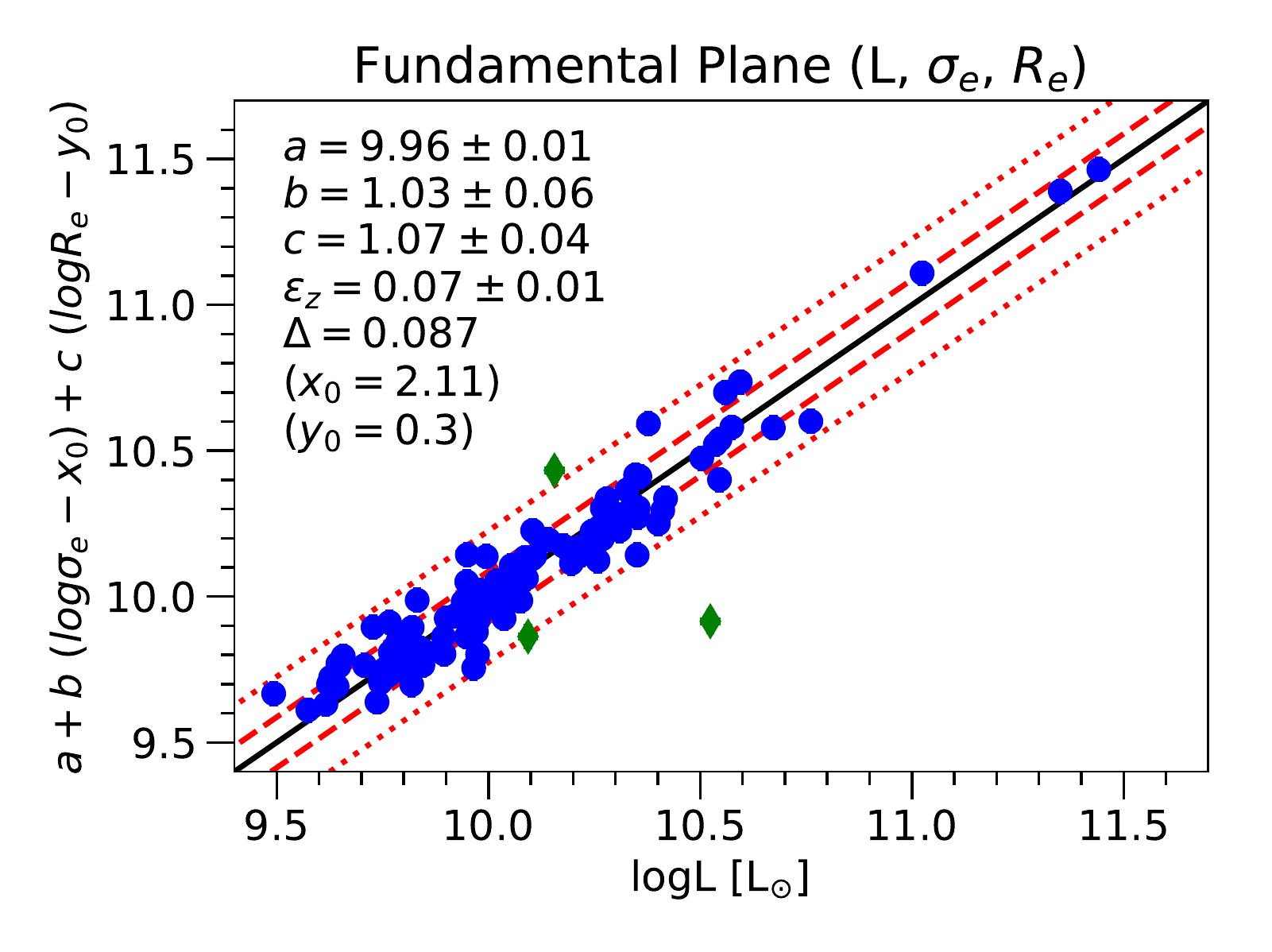}
	\caption[Fundamental Plane]{The Fundamental Plane for our galaxies in the full sample. The solid black line is the best fit solution by \textsc{LTS\_PLANEFIT}, with the red dashed and red dotted lines representing the 1$\sigma$ and 2.6$\sigma$ errors. The best-fitting coefficients, based on \textsc{LTS\_PLANEFIT} fitting, are presented in the top left-hand corner of the plot; the coefficients \textit{a}, \textit{b}, \textit{c}, $x_0$ and $y_0$ are the best fit coefficients for the plane fit, while $\epsilon_y$ and $\Delta$ are the intrinsic scatter along the y-axis of the plot and the root-mean-square of the data along the best fit respectively. The blue points represent the galaxies fitted, while the green points represent galaxies that were clipped during the fitting process. As seen in numerous previous studies, the best-fitting coefficients for the plane are not consistent with those expected by the Virial equilibrium condition under the assumption of a constant M/L, i.e. \textit{b}=2 and \textit{c}=1.}
	\label{FP_full}
\end{figure}

\begin{figure}
	\centering    
	\includegraphics[width=0.9\columnwidth]{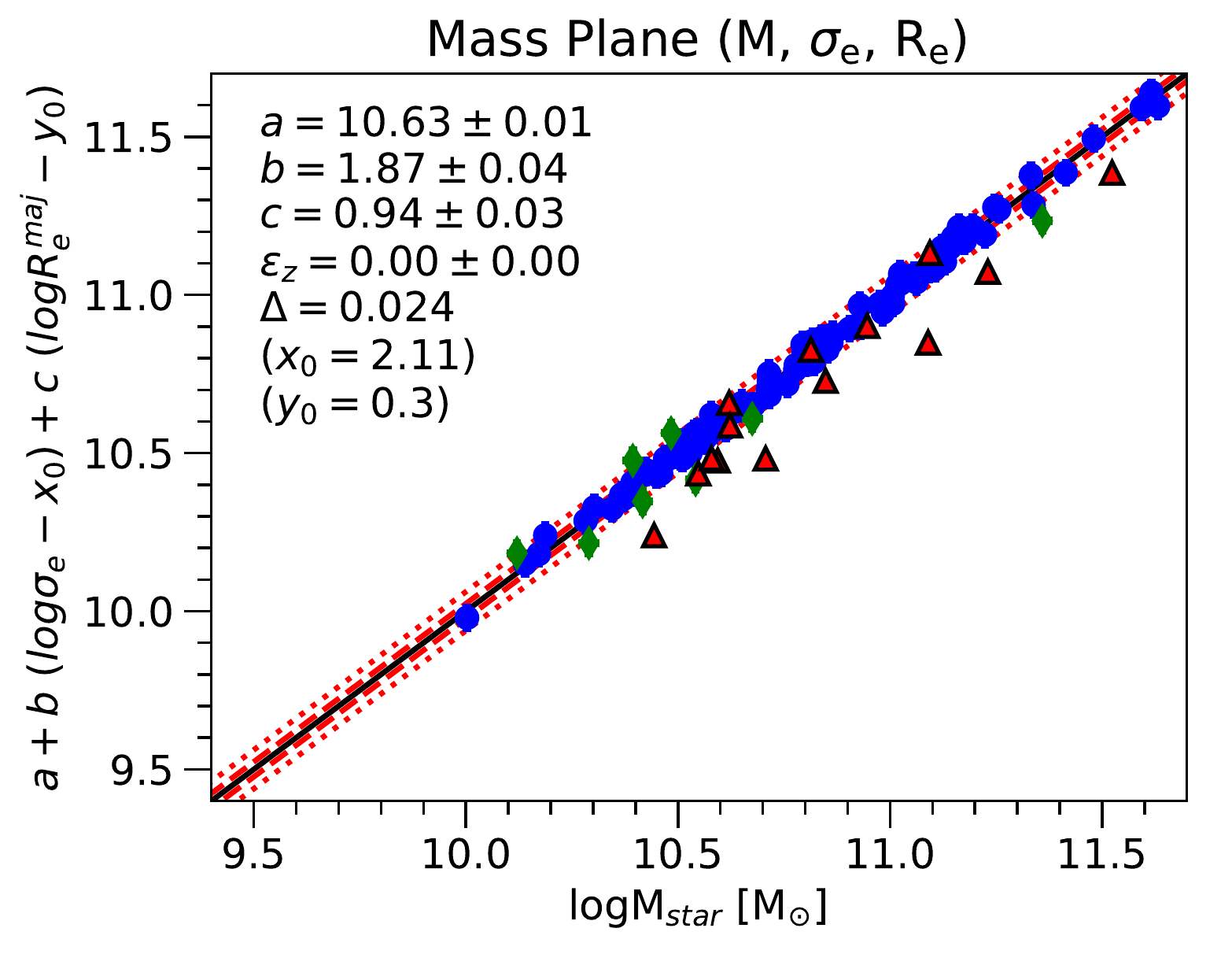}
	\caption{The mass plane for our galaxies in the full sample, with the best-fit solution derived using the \textsc{LTS\_PLANEFIT} and, similar to Fig.~\ref{FP_full}, its results are described along the top left corner of the plot. The mass plane is significantly closer to the predictions of the Virial equation than the Fundamental Plane. To complement the results of the full sample, we over-plot SAURON subset, with \textit{Qual} = 2, using results from their detailed dynamical models as red triangles. Like the full sample, the SAURON subset presents little scatter in the mass plane though these galaxies appear to be offset compared to the full sample.}
	\label{MP_full}
\end{figure}

\begin{figure}
	\centering
	\includegraphics[width=\columnwidth]{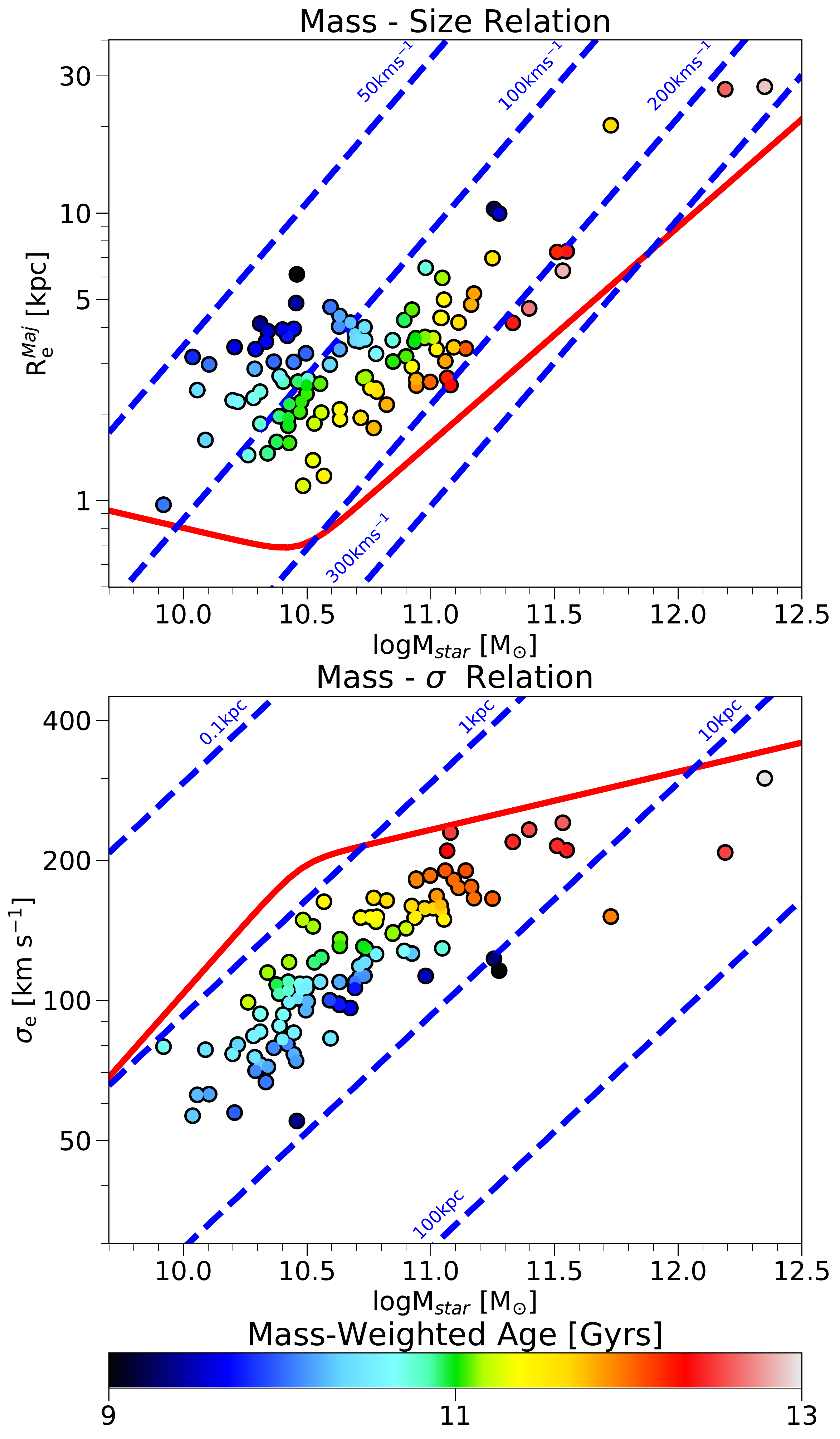}
	\caption[Scaling Relations of the Coma Cluster Galaxies]{Two projections of the Mass Plane (M$_{JAM}$, $\sigma_e$, R$_{e}^{Maj}$); The Mass-Size relation (top), and the Mass-$\sigma$ relation (bottom) of the galaxies. The circles in these plots denote the galaxies of the full sample, while their colour represents the LOESS smoothed mass-weighted age of the galaxies. The dashed blue lines represent lines of constant $\sigma$ and size in the Mass-Size and Faber-Jackson relation respectively. The galaxies in the Coma cluster appear to occupy similar regions of the Mass Plane as the \atl ~sample. However, the range of ages in the cluster sample is significantly lower than the ages observed in their field counterpart, providing evidence of environmental quenching of the cluster galaxies.}
	\label{ms_coma}
\end{figure}

\begin{figure}
	\centering
	\includegraphics[width=\columnwidth]{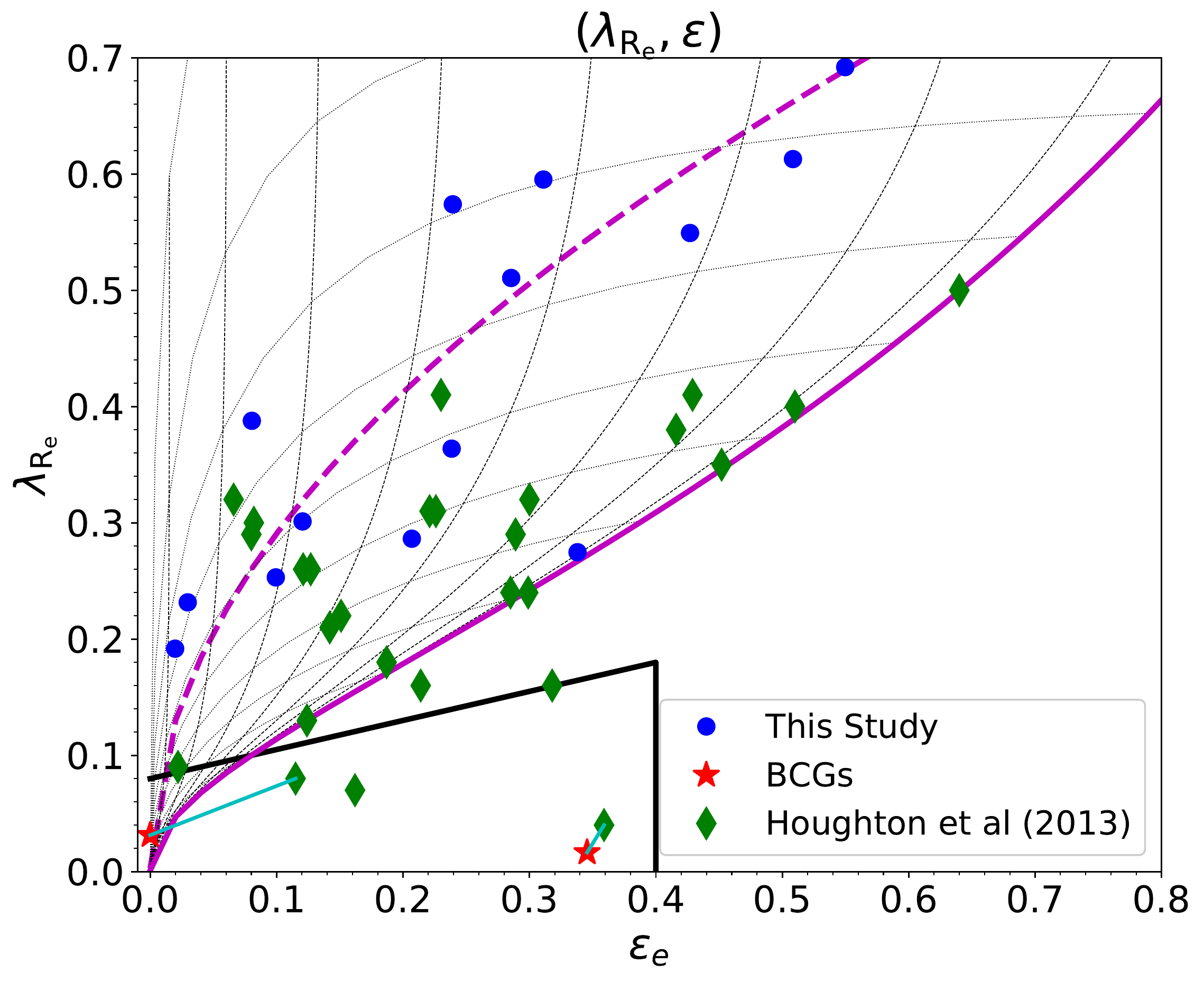}
	\caption[($\lambda_{\rm R_e}$, $\epsilon$)]{Here we present the ($\lambda_{\rm R_e}, \epsilon$) plot for our SAURON subset. The blue points represent the galaxies with \textit{Qual}$\geq$2, while the red stars represent our two BCGs. The green diamond points depict the galaxies in \citet{Houghton2013}. For the two BCGs, we connect the values measured by this study with those measured by \citet{Houghton2013} using solid-cyan lines. The dashed-magenta line represents the predicted position of an isotropic rotator at inclination 90\degree (edge-on), and the solid magenta line is the relation $\beta_{z} = 0.7 \times \epsilon$ of \citet{sauron10} between anisotropy and intrinsic ellipticity, which broadly describes the envelope of observed galaxies. The dashed black lines represent how the magenta line would evolve at different inclinations from edge-on. In contrast, the dotted-black line traces an anisotropic rotator in a range of inclinations for steps in intrinsic flatness. The plot demonstrates that galaxies in the Coma cluster tend to have lower stellar angular momentum, suggesting a larger bulge fraction than galaxies at low-density environments.}
	\label{anisotropy_coma}
\end{figure}

It is well known that spherical Jeans models suffer from the mass-anisotropy degeneracy \citep{BinneyMamon1982}. For this reason, one cannot determine both the mass profile (or BH mass) and anisotropy from these models without further assumptions. However, these days many BH models of core galaxies have been constructed, and it is not sensible to assume complete ignorance on the anisotropy in the inner regions of the galaxies. The orbits in the core regions are consistently found to be tangentially anisotropic (e.g. fig.~10 in \citealt{Gebhardt2003}) with the anisotropy restricted to the range $-0.5\la \beta\la 0.0$. When enforcing this constraint on the models, we found clear limits on the BH masses with crude errors of about a factor of two. Note that increasing $\beta$ always makes the sigma profiles steeper. For this reason, the upper limit on $\beta$ only affects models with excessively small BHs, while the lower bound effects models with the overly large BHs. Allowing for spatial variations in $\beta$, within the given ranges, would change the details of the model profiles but would not change our conclusions.

In Fig.~\ref{BCGs_SMBH}, we present the radially averaged velocity dispersion for our two BCGs and the best fitting spherical JAM models. Importantly, our best-fitting BH mass of 2.0$\times 10^{10} M_{\odot}$ for NGC4889 is entirely consistent with the best-fitting value by \citet{McConneletal2011} and \citet{McConnelletal2012} for the same galaxy, with independent data and models. The massive BH of 1.4$\times 10^{10} M_{\odot}$ for NGC4874 appears to confirm the finding that massive BCGs have larger BHs than expected for their velocity dispersion, due to their dry-mergers accretion history (e.g. sec.~6.7.2 of \citealt{KormendyHo2013}).

\subsection{(M/L)~--~$\sigma_{e}$ Relation}

Fig.~\ref{ml_vd_compare} presents the  \mljam~--~$\sigma_{e}$ relation for the Full sample and the SAURON subset of this study. In the plot we compare the two quantities using the \textsc{LTS\_LINEFIT} method\textsuperscript{\ref{noteMCsoftware}} \citepalias{atlas3d15} which combines the Least Trimmed Squares robust technique of \citet{rousseeuw2006computing} into a least-squares fitting algorithm which allows for errors in both variables and intrinsic scatter. The $\sigma_e$ ~presented in this study has been corrected, when necessary, to an aperture of 1\re ~using eq.~(1) of \citet{cappellari2006}. Previous studies indicated these quantities are tightly correlated and that they exhibit minimal variations within intermediate redshifts and different environment \citep[e.g. :][]{cappellari2006, vanderMarel2007}. Hence the \mljam~--~$\sigma_{e}$ relation is an ideal test for the accuracy of the dynamical analysis conducted in this study, and the consistency of the results between the two datasets. 

The relation derived for the full sample and the galaxies in the SAURON subset with \textit{Qual} = 2, are consistent within their $1\sigma$ errors, establishing the reliability of the dynamical models of the full sample. In the plot, the solid green line represents the \mljam~--~$\sigma_{e}$ relation observed in the global \atl ~sample. The consistency of the slopes between the Coma galaxies and \atl ~is encouraging and is as expected, given that both samples should be dominated by fast rotators. Furthermore Fig.~\ref{ml_vd_compare} also demonstrates the low RMS scatter ($\Delta=0.07$ or 17\%) observed in the relationship compared to the 28\% ($\Delta=0.11$) observed for the full sample of \atl ~and fig.~17 in \citetalias{Cappellari2016_review}. The reduction in scatter is due to the selection of the Coma cluster for this analysis, where the relative distance uncertainty between galaxies is significantly lowered. Hence the RMS scatter of the relationship is comparable to the $\Delta=0.058$ seen for the Virgo cluster galaxies in fig.~15 of \citetalias{atlas3d15} where the distances to the galaxies are well established. 

As mentioned in Section.~\ref{subsec:dm}, early-type galaxies have gradients in their stellar population properties, such as their stellar M/L. Hence we investigated the effect of these gradients by comparing our measured \mljam~ and \mlam with those from the dynamical models with a radial gradient (see Section~\ref{subsec:dm}). We find that in our extreme gradient scenario the measured \mljam~ and \mlam systematically increases by less than the random uncertainty in our measured values, i.e. $\lesssim$ 10\%, and hence is negligible.

\subsection{IMF-$\sigma$ Relation}

Fig.~\ref{imf_norm} presents the IMF normalization--$\sigma_{e}$ relation for the galaxies in the Coma galaxies. The plot also presents the relation derived by \citet{Posackietal2015_SALCS_IMF}, where the authors extend the dynamical determinations for the \atl ~sample with 55 massive ETGs of the SLAC survey \citep{slacs1} using lensing masses by \citet{Treuetal2010_SLACS}. The extended sample of \citet{Posackietal2015_SALCS_IMF} derives the trend for an extended velocity dispersion range and further strengthens the result of ATLAS$^{\rm 3D}$. The agreement between our study and that of \citet{Posackietal2015_SALCS_IMF} lends confidence to the analysis conducted in this study.

The plot demonstrates a low observed RMS scatter in the relationship between the two quantities. The best fit results of the relation, using abundance matching models to account for dark matter (top panel), has an observed RMS scatter of just 0.088 dex (23\%), compared to the relation found in \citetalias{atlas3d20} and \citet{Posackietal2015_SALCS_IMF} where both authors calculate a scatter of 0.12 dex (32\%). As seen in the previous section, this reduction is one of the outcomes we hoped to achieve through this study and caused by the significantly reduced relative uncertainty of the distances to the galaxies which has been a source of considerable error in the previous such studies. 

In the bottom panel, we reproduce the IMF normalization~--~$\sigma_e$ relation using the results of the self-consistent model for the galaxies. Together, the two panels demonstrate the robustness of the observed IMF normalization~--~$\sigma_e$ relation in the Coma cluster to the assumption of dark matter for the galaxies. 

A potential source of systematic in this analysis could be the inclusion of IMF sensitive features during the stellar population modelling of the galaxies. To account for the role of these features, we redid our stellar population analysis while masking the following IMF sensitive features: NaD [5875-5910\AA], TiO1 [5935-6000\AA], TiO2[6185-6275\AA], CaH1 [6355-6410\AA] and CaH2 [6770-6910\AA]. The resulting IMF-$\sigma_e$ relation was found to be consistent within one standard deviation of the results presented in Fig.~\ref{imf_norm}.

In addition to the effect of IMF sensitive features, gradients in stellar populations can have a systematic effect on this relation. In Sec.~\ref{subsec:dm} we demonstrate that the systematic effect due to strong gradients in M/L of ETGs is within the uncertainties on the measured dynamical stellar M/L. Additionally, \citet{Bernardietal2018} suggest that stellar M/L measured through dynamical techniques are more strongly affected by M/L gradients than those measured using stellar population modelling techniques. Hence the effect of M/L gradients on the measured IMF-$\sigma$ relation is unlikely to be larger than the uncertainties presented in the relationship here. However, the test on the effect of M/L gradients is simplistic, and a complete understanding should take into account the variation of the gradient with other galaxy properties \citep{DominguezSanchezetal2019}. Such an analysis is beyond the scope of this study.

When comparing the relation found in the Coma cluster with that seen for the global \atl ~+ SLACS galaxies, represented by the solid green line in the plots, an apparent offset is observed. This offset is likely the result of different SSP models being used, e.g. the stellar population modelling by \atl ~includes SSP models up to \textasciitilde{17Gyrs} while our models only extend out to \textasciitilde{14Gyrs}. 

\subsection{Scaling Relations}

Galaxies are known to follow tight correlations between their global parameters and hence these relations provide insight into the processes that affect galaxy evolution. Fig.~\ref{FP_full} presents the Fundamental plane for the full sample of our study. The best-fitting plane for the observed parameters has been determined by the \textsc{LTS\_PLANEFIT} code, the plane fitting variant of the previously used {\textsc LTS\_LINEFIT} \citepalias{atlas3d15}.

In Fig.~\ref{FP_full}, we present the Fundamental Plane of the galaxies along with the best fit solution and its one standard deviation (68\%) and 2.6 standard deviations (99\%). Comparing this to the Fundamental plane derived by \citetalias{atlas3d15}, we find the coefficients of the Fundamental Planes are consistent, but only within 3-$\sigma$ limit. This is expected given the near-identical selection and analysis methodology use between this study and \atl.

To compare our results with results by previous studies we fit the classic Fundamental plane, a plane in galaxy size (R$_{\rm e}$), central velocity dispersion ($\sigma_{\rm e}$) and effective surface brightness ($\Sigma_{\rm e}$), using the \textsc{LTS\_PLANEFIT}. We find our best-fit classic Fundamental plane to be log $R_{\rm e}$ = 1.05($\pm$ 0.06)log $\sigma_{\rm e}$ - 1.03($\pm$ 0.04)log $\Sigma_{\rm e}$ + c, with a RMS scatter of 0.093. Comparing this relation to the classic study by \citet{Jorgensenetal1996} of the Fundamental plane for E and S0 galaxies in clusters, we find that the coefficients diverge significantly. A similar inconsistency is observed by \citetalias{atlas3d15} when comparing their derived Fundamental plane to that measured by classical studies like \citet{Dressleretal1987, DjorgovskiDavis1987, GibbonsFruchterBothun2001, Bernardietal2003}, likely due to differences in sample selection. This is consistent with results by \citet{DOnofrioetal2008}, \citet{Gargiuloetal2009} and \citet{HydeBernardi2009} who demonstrated that differences in sample selection can significantly bias the measured Fundamental plane. In contrast, we find that our measured Fundamental plane is consistent with that measured by \citet{SAMI_FP_2015} using SAMI data for three low-redshift clusters.

With the derived stellar M/L from our generated dynamical models, using the abundance-matched models, and the measured \remaj, we transform the Fundamental plane into the mass plane for the galaxies in Fig.~\ref{MP_full}. We note that in this presentation of the Mass plane, we use the \remaj ~measured from the photometry of the galaxies and not the half-mass radius. This is because the size parameter of the virial relation varies with the light profile of the galaxies (at fixed mass density profile) and the mass density profile (at fixed light profile). This suggests that neither the half-light radius nor the half-mass radius fully-encapsulates the size parameter in the virial relation. However, due to the relatively stronger dependence on assumptions and difficulty in practical measurement of the half-mass radius, we prefer to present our Mass plane using \remaj.

To test the reliability of the Mass plane generated using the simple dynamical models of the Full sample, we have over-plotted the results from the SAURON subset in Fig.~\ref{MP_full}. The SAURON subset appear to be generally lower than the best-fit relation due to the fact a slight offset observed when comparing the derived velocity dispersion using SDSS and our SAURON data, with aperture correction and within a common wavelength range. The cause of this offset is unknown. The coefficients of our best-fit Mass plane are consistent with the previous result by \citetalias{atlas3d15} and \citealt{SAMI_FP_2015}, using detailed dynamical modelling, and by \citet{Augeretal2010} using strong lensing to determine the mass of the galaxies.

Given that the mass plane derived for the galaxies in the full sample is close to the Virial relation, studying this plane from other orientations contains information on the evolution of these galaxies. In Fig.~\ref{ms_coma}, we present two projections of the mass plane of the Coma galaxies, the mass-size and mass-$\sigma_e$ relations, to study their stellar populations. The size of the galaxy in the mass-size plot is represented by major axis of the isophote containing half of the galaxy. This quantity is preferred over the more standard effective radii due to its robustness against the inclination \citep{Hopkins2010}. The coloured points in the plots represent the LOESS-smoothed trends in the regularized mass-weighted ages of the galaxies. The solid red line in both plots cuts out the Zone of Exclusion (ZoE) of the plane, beyond which the probability of finding a galaxy drops significantly in the \atl ~sample \citepalias{atlas3d20}.  

From the figure it can be seen that galaxy ages follow lines of constant velocity dispersion, consistent with results seen in \atl (\citetalias{atlas3d20}; \citealt{ATLAS3D30}), MaNGA \citep{Lietal2018} and SAMI \citep{Scottetal2017}. Moreover, galaxies with masses M$_{\rm JAM}< 10^{11.5}$M$_{\odot}$ occupy the region in the scaling relations which are occupied by fast rotator galaxies in the low-density environment, suggesting that these galaxies may be fast rotators as well. This idea is supported by the kinematic maps observed for galaxies in the SAURON subset. When contrasting these fast rotators with those in the \atl ~sample, it would appear that the cluster galaxies are significantly older than their field counterpart, and in general, these cluster galaxies exhibit a smaller range in stellar ages than \atl. This result is consistent with the paradigm of galaxy evolution described by \citetalias{Cappellari2016_review}, wherein field fast rotators falling into cluster environments are unable to accrete gas, due to their relatively high speeds, and could lose gas to stripping. Due to this, cluster fast rotators are quenched and hence contain older stellar populations compared to their field counterparts.

\subsection{The ($\lambda_{\rm R_e}$, $\epsilon$) Diagram}

Building on the analysis of the spatially resolved stellar kinematics and photometry, we present in Fig.~\ref{anisotropy_coma} the ($\lambda_{\rm R_e}$, $\epsilon$) diagram for our galaxies in the Coma cluster \citep{Emsellem2007, sauron10, Jesseitetal2009}.

The parameter $\lambda_{\rm R_e}$ was measured using the following equation from \citet{SAURON9};

\begin{equation}
\lambda_{\rm R} \equiv \frac{\left< R|V| \right>}{\left< R\sqrt{V^2 + \sigma^2} \right>} = \frac{\sum_{n=1}^{N} F_n R_n |V_n|}{\sum_{n=1}^{N} F_n R_n \sqrt{V^2_n + \sigma^2_n}},
\end{equation}

\noindent where $F_{\rm n}$, $R_{\rm n}$, $V_{\rm n}$, and $\sigma_{\rm n}$ are the flux, radius from the center, velocity, and velocity dispersion of an individual spaxel. The summation is performed on all spaxels within the ellipse of containing half of the light as measured from the MGE models for the galaxy photometry.  For spaxels that do not contain the minimum S/N to measure the kinematics reliably, the kinematics of the Voronoi bin to which they were allocated in Section~\ref{sauron_kin} were assigned. For our two BCGs, the IFU coverage did not extend out to \remaj, and hence we present their specific angular momentum as measured within the field of view of the IFU.

In Fig.~\ref{anisotropy_coma}, we observe that for our SAURON subset of galaxies, the $\lambda_{\rm R_e}$ does not exceed a value of more than \textasciitilde{0.7}. This result is consistent with other measurements of $\lambda_{\rm R_e}$ in the Coma cluster by \citet{Houghton2013} where authors present the ($\lambda$-$\epsilon$) diagram for 27 galaxies in the Coma cluster. Comparing this to fig.~6 of \citet{Graham2018}, where the authors used the MaNGA IFS survey \citep{MANGA} data to measure the $\lambda_{\rm R_e}$ for \textasciitilde{2,300} galaxies in relatively low-density environments on average, the peak of stellar angular momentum in Coma cluster is significantly lower than \textasciitilde{0.9} seen in MaNGA. This disagreement is to be expected as the Morphology-Density relation \citep{Dressler1980} implies that the high density environment of Coma cluster would be dominated by ETGs with large bulge fractions which in turn are known to correlate with lower $\lambda_{\rm R_e}$ \citepalias[see section~3.6.3 of][]{Cappellari2016_review}.

In \citet{Houghton2013} the authors demonstrate the presence of three slow rotators in the cluster; NGC4889, NGC4874, and NGC4860. Our sample of SAURON observed galaxies does not add any more to this list, consistent with the estimate by \citet{Cappellari2013ApJL} who claim that it is unlikely for the cluster to contain any more. This result supports the hierarchical formation history for slow rotators presented by \citetalias{Cappellari2016_review} wherein central galaxies in centre of their groups tend to be the progenitors of the observed massive slow rotators. When groups/clusters merge their central galaxies also merge owing to their relatively high masses and low velocities with respect to their halos and form massive slow rotators in the halo centre. Due to this, the number fraction of slow rotators tend to remain constant or reduce as group and clusters grow to form massive galaxy clusters such as the Coma cluster.

\section{Summary}

We have studied a sample of 148 galaxies, selected from a magnitude-limited catalogue of 161 galaxies in the Coma cluster, that have single-fibre spectra and photometry available from SDSS. From this data, we have derived the stellar kinematics, parametrized the surface brightness profile of the galaxies, and created dynamical and stellar population models for the galaxies. Moreover, we obtained SAURON IFU observations for 53 galaxies of the sample of 148, of which we successfully created reliable dynamical models for 16, including the two BCGs. For the two BCGs, we improved the quality of the dynamical models by using high-resolution \textit{HST/ACS} observations which provide better data to model their light and stellar mass distribution. These detailed dynamical models robustly account for the dark matter content in the central regions of the galaxies by using two independent models for their total mass distribution, and we determine the stellar M/L of the galaxies using these dynamical models. Also, we use the full wavelength coverage of the SDSS spectrum to model the stellar populations of the galaxies using a non-parametric star formation history and use these models to determine the stellar M/L of the galaxies and mass-weighted stellar ages. 

Our analysis demonstrates that galaxies in the Coma cluster tend to have low dark matter fractions within a sphere of 1\re. This result is consistent with previous measures of the dark matter fractions in Coma cluster galaxies. Interestingly for our two BCGs we find a significant difference in the determined dark matter fractions for our two mass profiles with dark matter, the Abundance Matching model and the Power-Law model. Upon inspecting the predicted stellar kinematics for each of the two models, we find that the Abundance Matched model does not match the observed stellar kinematics as well as the Power-Law model. This inconsistency is expected, with recent results and suggests that the Abundance Matching may be predicting over-concentrated dark matter halos for BCGs. 

Using the dynamical models of the full sample, supported by the detailed dynamical models of the SAURON subset, we have presented the relationship between the IMF normalization of the galaxies and their central velocity dispersion. Owing to the cluster membership of the galaxies, all the galaxies can be approximated as being at the same distance of 100 Mpc, and at this distance the relative uncertainty of the distance to the individual galaxies is small. Hence, in Fig.~\ref{imf_norm}, the observed scatter in the relationship is significantly lower than that seen in fig.~13 in \citetalias{atlas3d20}. To our knowledge, this is the tightest relation between the IMF normalization and velocity dispersion published thus far. 

We also present the Fundamental Plane and the Mass Plane for the Coma cluster galaxies. As expected, we see a significant reduction in the scatter in the latter and find that the coefficients of the galaxy parameters are closer to that predicted by virial equilibrium. When we study the other projections of the mass plane, such as the mass-size or mass-$\sigma$ projections, we demonstrate that the mass-weighted age of the galaxies in the cluster varies systematically in those projections. These results are similar to those seen in \citetalias{atlas3d20} where galaxies with M$_{\rm JAM}< 2 \times 10^{11}$M$_{\odot}$ are likely fast rotators as seen in the high-density environment. Despite the identical variation in the stellar ages of the galaxies through the scaling relations compared to the local sample, the galaxies in the Coma cluster tend to be smaller. This suggests that despite the effect of environment, the evolutionary processes of galaxies are similar to that of the local universe and/or these galaxies may evolve only passively after their in fall into the cluster.

Finally, we present the ($\lambda_{\rm R_e}, \epsilon$) plot for our SAURON subset. The plot presents a picture consistent with what is expected based on the known morphology-density relation for the galaxies and is consistent with that expected by a hierarchical formation scenario for galaxies, especially a formation channel through dry mergers of central galaxies for slow rotators. 

\section*{Acknowledgements}
\addcontentsline{toc}{section}{Acknowledgements}

SS acknowledges support from National Science Foundation of China (11721303, 11991052), the National Key R\&D Program of China (2016YFA0400702) and NSF AST-1517006. 
RLD acknowledges travel and computer grants from Christ Church, Oxford, and support from the Oxford Hintze Centre for Astrophysical Surveys, which is funded through generous support from the Hintze Family Charitable Foundation. 
The SAURON observations were obtained at the WHT, operated by the Isaac Newton Group in the Spanish Observatorio del Roque de los Muchachos of the Instituto de Astrofisica de Canarias. We would like to thank the WHT staff for their support during observing runs. Funding for the SDSS and SDSS-II was provided by the Alfred P. Sloan Foundation, the Participating Institutions, the National Science Foundation, the US Department of Energy, the National Aeronautics and Space Administration, the Japanese Monbukagakusho, the Max Planck Society and the Higher Education Funding Council for England. The SDSS was managed by the Astrophysical Research Consortium for the Participating Institutions. Based on observations made with the NASA/ESA Hubble Space Telescope, obtained from the data archive at the Space Telescope Science Institute. STScI is operated by the Association of Universities for Research in Astronomy, Inc. under NASA contract NAS 5-26555. 

\bibliographystyle{mnras}
\bibliography{Coma_manuscript}

\label{lastpage}

\appendix

\begin{landscape}
	\section{Table 1}
	\label{Table_sauron}
	\begin{table}
		\begin{threeparttable}
			\caption{Results of analysis of the SAURON subset of Coma cluster galaxies. The table in its entirety is available as a supplementary file on the journal website.}  
			\begin{tabular}{cccccccccccccccccc}
				\toprule
				\toprule
				Galaxy & RA & Dec & \re & \remaj & $\sigma_{\tiny e}$ & $M_{{\tiny r}}$ & M/L$_{{\tiny \rm JAM}}$ & M/L$_{{\tiny dyn}}^{{\tiny AM}}$ & $f_{{\tiny \rm AM}}$ & log($\rho_{{\tiny tot}}$) & $\gamma$ & \mlpl & $f_{{\tiny \rm PL}}$ & log($M_*$) & $\epsilon$ & $\lambda_{{\tiny \rm R_e}}$ & \textit{Qual} \\
				{} & {\tiny ($^{\circ}$)} & {\tiny ($^{\circ}$)} & {\tiny (")} & {\tiny (")} & {\tiny (km s$^{-1}$)} & {\tiny (mag)} & {\tiny ($M_{\odot}/L_{\odot}$)} & {\tiny ($M_{\odot}/L_{\odot}$)} & {} & {\tiny (M$_{\odot}$pc$^{-3}$)} & {\tiny} & {\tiny ($M_{\odot}/L_{\odot}$)} & {} & {\tiny ($M_{\odot}$)} & {} & {} & {} \\
			    {(1)} & {(2)} & {(3)} & {(4)} & {(5)} & {(6)} & {(7)} & {(8)} & {(9)} & {(10)} & {(11)} & {(12)} & {(13)} & {(14)} & {(15)} & {(16)} & {(17)} & {(18)} \\
				\midrule
				{IC3955} &   194.7751 &    27.9966 &   6.63 &   6.68 &  152.8 &  -20.70 &   5.80 &   5.85 &  0.15 &  -2.92 &  -2.13 &   5.10 &  0.15 &  10.90 &  0.02 &  0.19 & 2 \\ 
				{IC3973} &   194.8783 &    27.8841 &   5.13 &   5.51 &  196.0 &  -20.97 &   6.00 &   5.59 &  0.09 &  -2.68 &  -2.12 &   4.41 &  0.24 &  10.99 & -- & -- & 1 \\ 
				{IC3976} &   194.8723 &    27.8502 &   3.46 &   5.20 &  263.3 &  -20.53 &  11.89 &  10.92 &  0.05 &  -2.79 &  -2.28 &   8.08 &  0.10 &  11.11 & -- & -- & 1 \\ 
				{IC3998} &   194.9451 &    27.9737 &   6.79 &   7.38 &  119.0 &  -20.48 &   5.43 &   4.82 &  0.17 &  -2.66 &  -1.78 &   3.50 &  0.45 &  10.73 & -- & -- & 1 \\ 
				{IC4011} &   195.0268 &    28.0040 &   5.03 &   5.26 &   88.2 &  -20.11 &   2.42 &   2.01 &  0.21 &  -4.15 &  -2.60 &   1.32 &  0.00 &  10.20 & -- & -- & 1 \\ 
				{IC4012} &   195.0335 &    28.0784 &   3.28 &   3.69 &  168.2 &  -20.34 &   5.40 &   5.21 &  0.05 &  -3.13 &  -2.24 &   3.49 &  0.07 &  10.71 &  0.21 &  0.27 & 2 \\ 
				{IC4021} &   195.0615 &    28.0412 &   3.78 &   4.00 &  142.5 &  -20.25 &   4.64 &   4.36 &  0.07 &  -3.07 &  -2.17 &   3.31 &  0.17 &  10.59 & -- & -- & 1 \\ 				
				\bottomrule
			\end{tabular}
		\begin{tablenotes}
			\small
			\item \textit{Notes}: Column (1) : Galaxy name. Column (2) : Right Ascension in degrees (J2000). Column (3) : Declination in degrees (J2000). Column(4) : Effective radii of the galaxies in arcsec. These quantities are derived from the MGE parametrization of the galaxy photometry and have been multiplied by 1.35 as per the offset observed in \atl. For more details, please refer to Section.~\ref{surf_brightness_parametrization}. Column (5) : Major axis of the half-light isopleth in arcsec. This is derived as per the prescription in \citetalias{atlas3d15} and has been multiplied by 1.35, as done for the effective radius. Column (6) : Log of the velocity dispersion measured within an aperture of 1\re in units of kms$^{-1}$. The error in the derived velocity dispersion is 0.042 dex or 10\%. Column (7) : Absolute magnitude of the galaxy derived from its $r$-band SDSS photometry. Column (8) : Log of the dynamical (M/L) of the galaxies derived using the Self-Consistent model. It is measured in units of (M$_{\odot}$/L$_{\odot}$), and has an error of 0.04 dex or 10\%. Column (9) : Log of the stellar (M/L) of the galaxies derived dynamically using the Abundance-Matching model. It is measured in units of (M$_{\odot}$/L$_{\odot}$) and similar errors as the dynamical \mljam. Column (10) : Dark matter fraction within 1\re~of the galaxies based on the Abundance Matched model. Column (11) : Log of the total density at the scale radius (20 kpc) of the best fitting power law density profile, as derived using the Power Law model. This value is measured in units of ($M_{\odot} pc^{-3}$). Column (12) : The slope, $\gamma$, of the best fitting power law density profile, as derived using the Power Law model. Column (13) : Log of the stellar (M/L) for the galaxies derived from the best fitting power law density profile, as derived using the Power Law model. Column (14) : Dark matter fraction within 1\re~of the galaxies based on the Power Law model. Column (15) : Log of the stellar mass of the galaxies in units of M$_{\odot}$, calculated using the galaxies absolute magnitude and stellar \mlam. Column(16) : Ellipticity of isophote with major-axis of \remaj. Column (17) : Proxy for angular momentum measured within isophote with major-axis of \remaj and ellipticity $\epsilon$. Column (18) : Quantity indicating the quality of the dynamical models based on the results of the Self-Consistent model.
		\end{tablenotes}
		\end{threeparttable}
		
	\end{table}
\end{landscape}

\clearpage

\section{Table 2}
\label{Table_full}

\begin{table}
	\begin{threeparttable}
		\caption{Results of analysis of the full sample of Coma cluster galaxies. The table in its entirety is available as a supplementary file on the journal website.}  
		\begin{tabular}{ccccccccccccc}
			\toprule
			\toprule
			Galaxy & RA & Dec & \re & \remaj & $\sigma$ & $M_{\tiny r}$ & M/L$_{{\tiny \rm JAM}}$ & M/L$_{{\tiny dyn}}^{{\tiny AM}}$ & M/L$_{\tiny pop}^{\tiny Salp}$ & Age & log($M_*$) & SAURON \\
			{} & {\tiny ($^{\circ}$)} & {\tiny ($^{\circ}$)} & {\tiny (")} & {\tiny (")} & {\tiny (km s$^{-1}$)} & {\tiny (mag)} & {\tiny ($M_{\odot}/L_{\odot}$)} & {\tiny ($M_{\odot}/L_{\odot}$)} & {\tiny ($M_{\odot}/L_{\odot}$)} & {\tiny (Gyrs)} & {\tiny ($M_{\odot}$)} & {} \\
			{(1)} & {(2)} & {(3)} & {(4)} & {(5)} & {(6)} & {(7)} & {(8)} & {(9)} & {(10)} & {(11)} & {(12)} & {(13)} \\
			\midrule				
		    IC3943 &   194.6515 &    28.1138 &   5.02 &   7.58 &  170.8 &  -20.81 &   5.70 &   5.68 &   5.86 & 11.42 &  10.93 & 0 \\
			IC3946 &   194.7029 &    27.8102 &   5.01 &   6.30 &  205.7 &  -21.04 &   5.61 &   5.76 &   5.95 & 12.05 &  11.03 & 0 \\
			IC3947 &   194.7170 &    27.7850 &   4.31 &   5.49 &  139.9 &  -20.41 &   4.66 &   4.53 &   5.38 & 11.39 &  10.68 & 0 \\
			IC3955 &   194.7751 &    27.9966 &   6.63 &   6.68 &  138.7 &  -20.70 &   3.99 &   3.70 &   5.06 &  9.95 &  10.71 & 1 \\
			IC3957 &   194.7811 &    27.7678 &   4.81 &   4.93 &  163.4 &  -20.39 &   5.35 &   5.28 &   6.11 & 12.35 &  10.74 & 0 \\
			IC3959 &   194.7842 &    27.7841 &   7.04 &   7.04 &  201.0 &  -21.08 &   5.83 &   5.79 &   6.36 & 12.68 &  11.05 & 0 \\
			IC3960 &   194.7830 &    27.8550 &   5.06 &   5.06 &  159.9 &  -20.47 &   4.95 &   4.82 &   6.42 & 12.53 &  10.73 & 0 \\
			\bottomrule
		\end{tabular}
	\begin{tablenotes}
		\small
		\item \textit{Notes}: Column (1) : Galaxy name. Column (2) : Right Ascension in degrees (J2000). Column (3) : Declination in degrees (J2000). Column(4) : Effective radii of the galaxies in arcsec. These quantities are derived from the MGE parametrization of the galaxy photometry and have been multiplied by 1.35 as per the offset observed in \atl. For more details, please refer to Section.~\ref{surf_brightness_parametrization}. Column (5) : Major axis of the half-light isophote in arcsec. This is derived as per the prescription in \citetalias{atlas3d15} and has been multiplied by 1.35, as done for the effective radius. Column (6) : Log of the velocity dispersion measured in km s$^{-1}$. The error in the derived velocity dispersion is 0.042 dex or 10\%. Column (7) : Absolute magnitude of the galaxy derived from its r-band SDSS photometry. Column (8) : Log of the dynamical (M/L) of the galaxies derived using the Self-Consistent model. It is measured in units of (M$_{\odot}$/L$_{\odot}$), and has an error of 0.04 dex or 10\%. Column (9) : Log of the stellar (M/L) of the galaxies derived dynamically using the Abundance-Matching model. It is measured in units of (M$_{\odot}$/L$_{\odot}$) and similar errors as the dynamical \mljam. Column (10) : Stellar Mass-to-Light ratio in units of (M$_{\odot}$/L$_{\odot}$). These are derived by the regularized mass-weighted fitting of the galaxy spectra with stellar population models, assuming a Salpeter IMF. Column (11) : Mass-weighted stellar age of the galaxies in Gyrs. Column (12) : Log of the stellar mass of the galaxies in units of M$_{\odot}$, calculated using the galaxies absolute magnitude and stellar \mlam. Column (13) : Quantity stating if the galaxy has SAURON observations (=1) or not (=0).
	\end{tablenotes}
	\end{threeparttable}
\end{table}

\end{document}